\documentclass[aps,prb,manuscript]{revtex4}
\usepackage{epsfig}
\usepackage{amsmath}
\begin{document}
\draft
\title{Field enhanced electron mobility by nonlinear phonon scattering of Dirac electrons
in semiconducting graphene nanoribbons}

\author{Danhong Huang,$^{1}$ Godfrey Gumbs,$^{2}$ and O. Roslyak$^{2}$}

\address{$^{1}$Air Force Research Laboratory, Space Vehicles
Directorate,\\
Kirtland Air Force Base, NM 87117, USA}
\address{$^{2}$Department of Physics and Astronomy, Hunter College of the
City University of New York, 695 Park Avenue, New York, NY 10065,
USA}

\date{\today}

\begin{abstract}
The calculated electron mobility for a graphene nanoribbon as a function of  applied
electric field has been found to have a large threshold field for entering a nonlinear
transport regime. This field depends on the lattice temperature, electron density, impurity
scattering strength, nanoribbon width and correlation length for the line-edge roughness.
An enhanced electron mobility beyond this threshold has been observed, which is related to
the initially-heated electrons in high energy states with a larger group velocity. However, this mobility
enhancement quickly reaches a maximum due to the Fermi
velocity in graphene and the dramatically increased phonon scattering. Super-linear and sub-linear
temperature dependence of mobility  seen in the linear and nonlinear transport regimes.
By analyzing the calculated non-equilibrium electron distribution function, this
difference is attributed separately to the results of sweeping electrons from the right Fermi
edge to the left one through the elastic scattering and moving electrons from low-energy states
to high-energy ones through field-induced electron heating. The threshold field is pushed up by a
decreased correlation length in the high field regime, and is further accompanied by a reduced
magnitude in the mobility enhancement. This implies an anomalous high-field increase of the line-edge
roughness scattering with decreasing correlation length due to the occupation of high-energy states
by field-induced electron heating.
\end{abstract}


\maketitle

\section{Introduction}
\label{sec1}

The engineering achievement of isolating graphene sheets\,\cite{art10,art11,art12}
from graphite has inspired many studies aimed at understanding basic underlying physics\,\cite{art10,neto}
as well as finding possible  applications to carbon-based electronics\,\cite{art13}.
The linear transport of charge carriers in a graphene layer, has received a considerable
amount of attention.\,\cite{art1,art2,art3,art4,art5,art7,art8} Recent reports on the
successful fabrications of ultra-fast graphene transistors\,\cite{art14} and
photodetector\,\cite{art15} have further pushe d this research frontier into the
fields of electronics and optoelectronics. However, similar studies of
linear transport in graphene nanoribbons (GNRs)has only been given relatively
little attention so far.\,\cite{art6,fang,art9}
\medskip

Early theoretical studies\,\cite{art7,fang} on electron transport in graphene
nanoribbons were restricted to the low-field limit, where a linearized Boltzmann
equation was solved within a relaxation-time approximation. In this paper, the non-equilibrium
distribution of electrons is calculated exactly by solving the Boltzmann equation
beyond the relaxation-time approximation for nonlinear electron transport in semiconducting
graphene nanoribbons. Enhanced electron mobility from initially-heated electrons in high energy states
is anticipated. An anomalous increase in the line-edge roughness scattering for
large electric fields is obtained with decreasing roughness correlation length due
to the occupation of high-energy states by field-induced electron heating.
Although the numerical results in this paper are given only for semiconducting graphene nanoribbons,
the derived formalism here can also be applied to metallic graphene nanoribbons by formally taking the limit $W\rightarrow\infty$.
The semi-classical Boltzmann transport equation is expected to be applicable to the diffusive band-transport regime
with relative smooth edges for graphene nanoribbons,
instead of the hopping and tunneling between localized states with rough edges.
\medskip

The rest of the paper is organized as follows. In Sec.\ \ref{SEC:Model}, we exactly solve
the semi-classical Boltzmann equation for low-temperature electron transport in semiconducting
graphene nanoribbons by including impurity, line-edge roughness and phonon scattering
effects at a microscopic level. Based on the calculated non-equilibrium distribution
as a function of wave number along the ribbon, we present detailed numerical results
for electron mobility as a function of either the applied electric field or the lattice
temperature for various electron densities, impurity scattering strengths,
nanoribbon widths and correlation lengths for line-edge roughness.
The conclusions deduced from these results are briefly summarized in Sec.\ \ref{sec3}.

\section{Model and Theory}
\label{SEC:Model}

\begin{figure}[p]
\centering
\epsfig{file=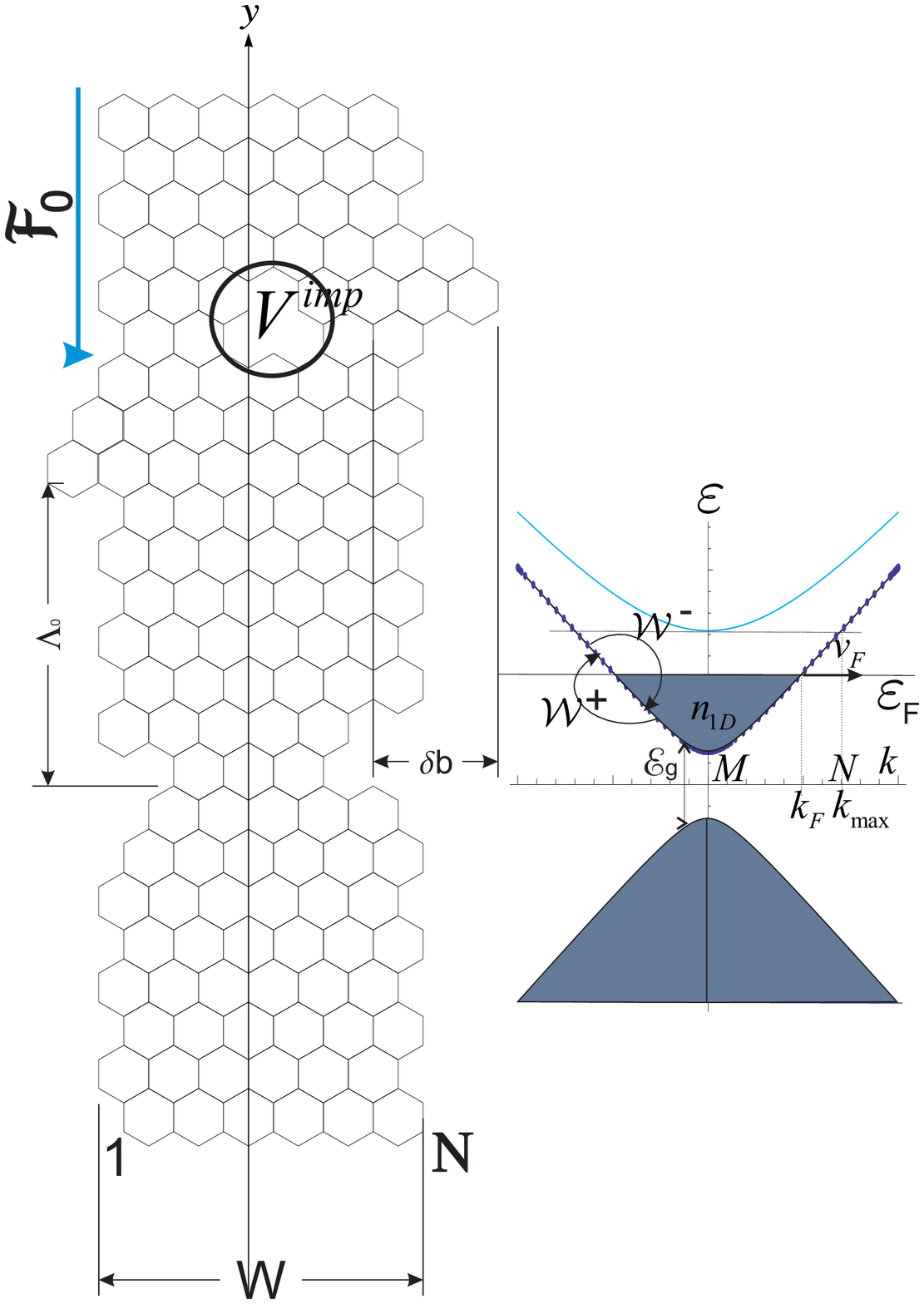,width=4.0in}
\caption{Schematic illustration of the model employed in our numerical calculations.
The parameters used in this figure are justified  in the text.}
\label{fig:HuangGraph}
\end{figure}

We restrict ourselves to  single subband transport. Low electron densities, moderate
temperatures, ionized impurities and line-edge roughness are assumed\,\cite{fang,huang1,huang2}.
Therefore, negligible effects of electron-electron\,\cite{pair},
optical phonon\,\cite{huang2}, inter-valley and volume-distributed impurity scattering\,\cite{fang}
will be ignored. Typical armchair nanoribbon (ANR) configuration is shown in Fig.\,\ref{fig:HuangGraph}.
The electron-like dispersion equation branch for ANR can be represented on a $k-$mesh as\,\cite{neto}

\begin{equation}
\label{EQ:DISPERSION}
\varepsilon_j = \hbar\nu_F\times\Bigg\{{\begin{array}{ll}
k_j &,\ \ \ \ \ \ \ \ \mbox{metallic}\\
\sqrt{k^2_j +\left({\pi/3W}\right)^2} &,\ \ \ \ \ \ \ \ \mbox{semiconducting}
\end{array}}\ .
\end{equation}
Here, $\nu_F=10^8$\,cm/s is the Fermi velocity in graphene. The wave numbers
$k_j=\left[{j-(N+1)/2}\right]\,\delta k$ are given on the discrete mesh
$j=1,\,2,\,\ldots,\,N$ for large odd integer $N$.
$\delta k=2\,k_{\rm max}/(N-1)$ is a small mesh spacing. $k_{\rm max}$ is
chosen so that scattering induced population of higher elctron-like branches may be neglected
(see Fig.\,\ref{fig:HuangGraph}). The minimum in the dispersion curve corresponds to the
central mesh point $j=M=(N+1)/2$. Also, $W=(\mathcal{N}+1)\,a_0/2$ is the width of
the ribbon expressed in units of the size of the graphene unit cell $a_0=2.6$\,\AA\ and
number of carbon atoms across the ribbon $\mathcal{N}$. According to the dispersion relation
in Eq.\,(\ref{EQ:DISPERSION}), the electron group velocity $v_j$ for metallic nanoribbons on the first subband is
determined by  the Fermi velocity $\nu_F$, while for semiconducting ANRs,
it assumes the form $v_j=\nu_F\left({\hbar\nu_{F}k_j/\varepsilon_j}\right)$.
We shall also assume that the electron-like branch is filled up to $\lvert{k_j}\rvert=k_F$ at zero temperature with the Fermi
wave number and energy given by $k_F=\pi n_{1D}/2$ and $\varepsilon_F=\varepsilon(k_F)$, which can be applied to both metallic and semiconducting graphene nanoribbons
by choosing the zero energy at the subband edge. For
a chosen temperature $T$ and chemical potential $\mu_0$, the linear electron density in
ANR follows from $n_{1D}=\delta k/\pi\,\sum\limits_{j=1}^N\,\left[{\exp{((\varepsilon_j-\mu_0)/\mathrm{k}_B T)}+1}\right]^{-1}$.
\medskip

The wave function $\Psi_j(x,\,y)$ corresponding to the energy in Eq.\,(\ref{EQ:DISPERSION})
must satisfy hard-wall boundary conditions\,\cite{fertig} $\Psi_j(0,\,y)=\Psi_j(W,\,y)=0$. This can
be assured by choosing the wave function as a mixture of those at $\mathbf{K}=\left({2\pi/3\,a_0,\,2\pi/\sqrt{3}\,a_0}\right)$ and
$\mathbf{K}^\prime=\left({-2\pi/3\,a_0,\,2\pi/\sqrt{3}\,a_0}\right)$ points as\,\cite{neto}

\begin{gather}
\label{EQ:WAVEFUNCTION}
\Psi_j(x,\,y)=\frac{1}{\sqrt{2}}\left[{\psi_j (x,\,y)-\psi^\prime_j(x,\,y)}\right]\ ,\\
\notag
\\
\left\{
\begin{array}{ll}
\psi_{j}(x,\,y)=\sqrt{\frac{1}{2LW}}\,e^{ik_jy}\,\left[{\begin{array}{c} 1\\ e^{i\phi_{j}}\end{array}}\right]\,e^{i(2\pi/3a_0-\kappa)x}
&\ \ \ \ \mbox{at $\mathbf{K}$ point}\\
\\
\psi^\prime_{j}(x,\,y)=\sqrt{\frac{1}{2LW}}\,e^{ik_jy}\,\left[{\begin{array}{c} 1\\ -e^{-i\phi_{j}}\end{array}}\right]\,e^{-i(2\pi/3a_0-\kappa)x}
&\ \ \ \ \mbox{at $\mathbf{K}^\prime$ point}
\end{array}\right.\ .
\end{gather}
Here, $L$ is the quantization length of the ribbon. For semiconducting ANR
$\kappa=\pi/3W\ll 2\pi/3a_0$ is the quanta of the transverse wave vector.
$\phi_j=\tan^{-1}\left({k_j/\kappa}\right)$ is the phase separation between the two graphene
sublattices. For metallic type of = ribbon, we have $\kappa=0$ and the phase assumes only $\pm\pi/2$ values.
\medskip

Given the wave function in Eq.\,(\ref{EQ:WAVEFUNCTION}) and neglecting effects due to inter-valley scattering\,\cite{fn1}
one can calculate the scattering from any given potential $V(x,\,y)$.
The impurity and phonon induced inter-valley scattering has been ignored based on the assumption that the required momentum transfer
is relative large in comparison with the low-temperature phonon energies as well as the suppressed effective scattering cross section of both
volume and surface impurities for very large value $|\mathbf{K}-\mathbf{K}^\prime|$. Corresponding interaction matrix elements are

\begin{equation}
V_{i,\,j}=\int\limits_{0}^{W} dx\int\limits^{\infty}_{-\infty} dy\,\Psi^\ast_i(x,\,y)\,V(x,\,y)\,\Psi_j(x,\,y)
=\frac{1}{2}\,\int\limits_{0}^{W} dx\int\limits^{\infty}_{-\infty} dy\,\left({\psi^\ast_i\,V\,\psi_j+\psi^{\prime\,\ast}_i\,V\,\psi^\prime_j}\right)\ .
\end{equation}
Our model accounts for the net scattering potential made from the three components
$V=V^{\rm AL}+V^{\rm LER}+V^{\rm imp}$.
\par

The longitudinal mode considered in this paper induces higher deformation potential than the out-of-plane flexural modes, and we neglect
the transverse flexural modes here as a leading-order approximation. In this approximation,
the inelastic scattering is provided only by longitudinal acoustic phonons.
The perturbation potential induced by them is given by\,\cite{fang,heat}

\begin{equation}
\label{EQ:ALPOTENTIAL}
V^{\rm AL}(y)=\sqrt{\frac{n^{\pm}\hbar}{2\rho LW\omega_{AL}}}\,D_{AL}q_y\,e^{iq_yy}\ ,
\end{equation}
where the phonon distributions are
$n^{-} = \left[{\exp{(\hbar\omega_{AL}/\mathrm{k}_B T)-1}}\right]^{-1}$, $n^{+}=1+n^{-}$,
$T$ is the temperature,
$\omega_{AL}=c_s\,q_y$ is the phonon frequency,
$D_{AL}\sim 16$\,eV is the deformation potential,
$\rho\sim 7.6\times 10^{-8}$\,g/cm$^2$ and
$c_s\sim 2\times 10^6$\,cm/s are the mass density and sound velocity in graphene.
Also, one has to assure  momentum conservation during scattering, i.e., $q_y=k_i-k_j$.
\medskip

Elastic scattering is due to the roughness of the ribbon edges as well as in-plane
charged impurities. To model the first  we shall assume that the width
of the ribbon has the form $W(y)=W+\delta W (y)$, where
the edge-roughness is assumed to satisfy Gaussian correlation function
$\langle\delta W(y)\,\delta W(y+\Delta y)\rangle=\delta b^2\,\exp[-(\Delta y/\Lambda_0)^2]$
with $\delta b\sim 5$\,\AA\ being the amplitude and $\Lambda_0\sim 50-200$\,\AA\ being the correlation length of the roughness.
The second mechanism is given by impurities located at $(x_0,\,0,\,0)$ and distributed with a sheet density $n_{\rm 2D}$.
Each of the point impurities provide the scattering Coulomb potential.
Since the momentum difference between two valleys is
rather large, $\vert\mathbf{K}-\mathbf{K}^\prime\vert=4\pi/3a$, only short-range
impurities (such as topological defects) with a range smaller than the lattice
constant causes inter-valley scattering.
In contrast, for wide enough ribbons with uniformly distributed long-range impurities
(such as charged defects) the scattering processes is restricted
to intra-valley scattering\,\cite{ando,wakabayashi}.
Here we neglect the short range scatterers in favor of the charged impurities thus making
the scattering matrix diagonal with all elements given by Eq.\,(7).
Such simplification has been invoked in order to keep the conventional part of the Boltzmann transport equation as simple as possible
and focus on the nonlinear counterpart.
Corresponding perturbations for roughness and impurity scattering are given by

\begin{eqnarray}
\label{EQ:LERPOTENTIAL}
V^{\rm LER}(y) & = & \frac{\delta W (y)}{3W^2}\,\pi\hbar\nu_F\ ,\\
\notag
\\
\label{EQ:IMPPOTENTIAL}
V^{\rm imp}(x,\,y) & = & \frac{e^2}{4\pi\epsilon_0\epsilon_r\sqrt{(x-x_0)^2 + y^2}}\ ,
\end{eqnarray}
where $\epsilon_r$ is the average dielectric constant of the host.
These potentials give rise to the net elastic scattering rate $\hbar/\tau_j$ on the $k-$mesh

\begin{gather}
\label{EQ:ELASTICSCATTERINRATE}
\frac{1}{\tau_j} = \frac{1}{\tau^{imp}_j}+\frac{1}{\tau^{LER}_j}\ ,\\
\notag
\\
\label{EQ:elasticrate}
\left\{
\begin{array}{ll}
\left(\tau^{imp}_j\right)^{-1}=\gamma_0\,\left(\frac{v_{\rm F}}{|v_j|}\right)\,\left[1+\cos(2\phi_j)\right]\\
\left(\tau^{LER}_j\right)^{-1}=\gamma_1 \left({\frac{\nu_F}{|v_j|}}\right)\frac{1}{1+4k_j^2\Lambda_0^2}\,\left[1+\cos(2\phi_j)\right]
\end{array}\right.\ ,
\end{gather}
where $\gamma_0$ denotes the impurity scattering rate at the Fermi edges
and $\gamma_1 = 2\left(\frac{\pi\nu_F\delta b}{3W^2}\right)^2\frac{\Lambda_0}{\nu_F}$
is the scattering rate due to edge roughens.
Here, we assume that the impurities lie within the graphene layer. The scattering rate will be
modified for charged impurities in a substrate, which can be easily seen from the effective scattering cross section ${\cal S}(k_y,\,W)$ defined
in Eq.\,(28) of Ref.\,[\onlinecite{fang}]. Although this effect can be formally taken into account by varying the value of $\gamma_0$ in Eq.\,\eqref{EQ:elasticrate},
the leading scattering mechanism is still provided by in-plane impurities.
All the scattering potentials are screened
by free carriers. Since we are limited to a single subband, the screening is described by
the dielectric function. The inelastic scattering is shielded by the static Thomas-Fermi
dielectric function in its general form\,\cite{fang,fertig1}
$\epsilon_{TF}(\lvert{k_{j^\prime}-k_j}\rvert)$. The screening of elastic scattering potentials is
given by $\epsilon_{TF}\approx 1+(e^2/\pi^2\epsilon_0\epsilon_r\hbar\nu_F)$ under the metallic limit ($2k_F W\gg 1$) with
$\epsilon_r\approx 3.9$.
We employ the static screening to both impurity and phonon scattering with electrons for the system with a relative large damping rate and small Fermi energy.
The static screening tends to overestimate the scattering effect.\,\cite{damp}
The dynamic screening effect on the electron scattering has been quantitatively evaluated within the RPA in a similar quantum-wire system,\,\cite{damp}
and it becomes important whenever the damping energy is much smaller than the Fermi energy.
The Fermi-golden rule has been employed to treated the leading contributions of both elastic and inelastic scattering.
The multiple scattering effects are neglected due to low temperatures, low impurity densities and small line-edge roughness.
Under application of a small electric field $\mathcal{F}_0$ along the GNR
(see Fig.\,\ref{fig:HuangGraph}) the stationary current $I=(e\delta k/\pi)\sum\limits_{j}\,f_j\,v_j$
establishes itself. It can be characterized by the electron mobility $\mu_e=I/en_{1D}{\cal F}_0$,
where we have introduced the carrier distribution function partitioned as
$f_j=f^{(0)}_j+g_j\approx f^{(0)}_j+b_j\,\tau^{TOT}_j$
with $f^{(0)}_j = \left\{1+\exp\left[(\varepsilon_j-\mu_c)/\mathrm{k}_B T\right]\right\}^{-1}$
being the equilibrium Fermi-Dirac function,
$\mu_c$ being the chemical potential of electrons,
and $b_j=ev_j\,(\partial f^{(0)}_j/\partial \varepsilon_j)\,\mathcal{F}_0$ being the drag.
Such approximation for the distribution function is known as relaxation time approximation and
encompasses all (elastic and inelastic) scattering mechanisms into $\tau^{TOT}_j$.
The consequence of such an approximation is the electric field independent mobility and conductivity.
However, by increasing $\mathcal{F}_0$ one may enter nonlinear regime when the conductivity
and mobility become a function of the applied electric field.
Studying of ANR in such nonlinear regime is the main subject of this work.
\medskip

Formally, the deviation from the equilibrium Fermi distribution under strong electric field
is described by the set of nonlinear reduced Boltzmann transport equations\,\cite{huang1,huang2}

\begin{equation}
\frac{dg^\prime_j(t)}{dt}=b_j-\sum_{j^{\,\prime}\neq M}\,a^\prime_{j,\,j^{\,\prime}}(t)\,g^\prime_{j^{\,\prime}}(t)\ ,
\label{e1}
\end{equation}
In this notation, $g^\prime_j(t) = g_j(t)-g_M (t)$ is the reduced form of the dynamical non-equilibrium
part\,\cite{fn2} of the electron distribution function. The reduced form accounts for
particle number conservation condition, i.e.,  $\sum \limits_{j=1}^N g_j(t) =0$.
The nonlinear reduced Boltzmann transport equations have been applied to study the electron dynamics in both quantum wires\,\cite{huang1}
and quantum-dot superlattices\,\cite{huang2}, where the full description on the derivation of the nonlinear reduced Boltzmann transport
equations can be found.
In the matrix Eq.\,(\ref{e1}), we introduced the matrix elements
$a^\prime_{j,\,j^{\,\prime}}(t)=a_{j,\,j^{\,\prime}}(t)-a_{j,\,M}(t)$
via its components

\begin{eqnarray}
a_{j,\,j^{\,\prime}}(t)&=&\delta_{j,\,j^{\,\prime}}\left[{\cal W}_j+{\cal W}_j^g(t)+
\frac{1-\delta_{j,\,(N+1)/2}}{2\tau_j}\right]-
\delta_{j+j^{\,\prime},\,N+1}\left[\frac{1-\delta_{j,\,(N+1)/2}}{2\tau_j}\right]
\nonumber\\
&-& {\cal W}_{j,\,j^{\,\prime}}-\frac{e{\cal F}_0}{2 \hbar\delta
k} \left({\delta_{j,j^\prime-1}-\delta_{j,j^\prime+1}}\right) \ .
\label{e7}
\end{eqnarray}
While the net elastic scattering rate is given by Eq.\,(\ref{EQ:ELASTICSCATTERINRATE}),
the total inelastic rate assumes the form $\mathcal{W}_j = 1/\tau_{j}^{AL} = \sum \limits_{j^\prime} \mathcal{W}_{j,j^\prime}$
with the scattering matrix given by

\begin{eqnarray}
\label{e14}
{\cal W}_{j,\,j^{\,\prime}} & = & \frac{L}{2\pi}\,\delta k \sum_\pm\,{\cal W}^\pm_{j,j^\prime}\,\left({n_{j,j^\prime}+f^{\pm}_{j^\prime}}\right)\\
W^{\pm}_{j,j^\prime} & = &
\theta(\pm\varepsilon_{j^{\,\prime}}\mp\varepsilon_j)\,\left[
\frac{D_{AL}^2|\varepsilon_{j^{\,\prime}}-\varepsilon_j|}{2\hbar^2c_s^3
\rho L W \epsilon^2_{\rm TF}(|k_{j^{\,\prime}}-k_j|)}\right]\,
\left[1+\cos(\phi_{j^{\,\prime}}-\phi_j)\right]\ .
\end{eqnarray}
We have $f^{-}_j=f^{(0)}_j$, $f^{+}_j=1-f^{(0)}_j$, $n_{j,j^\prime}
=N_0(|\varepsilon_{j^{\,\prime}}-\varepsilon_j|/
\hbar)$, $N_0(\omega_q)=[\exp(\hbar\omega_q/k_{\rm B}T)-1]^{-1}$ is the Bose function for
thermal equilibrium phonons.
Furthermore, the dynamical phonon scattering rate ${\cal W}_j^g(t)$ for nonlinear phonon scattering,
which is also responsible for the nonlinear electron transport and electron heating
due to its dependence on $g^\prime_{j^{\,\prime}}(t)$, is in the form of

\begin{eqnarray}
{\cal W}_j^g(t)=\frac{L}{2\pi}\,\delta k\,
\sum_{j^{\,\prime}\neq M}\,g^\prime_{j^{\,\prime}}(t) \left[{\cal W}^{+}_{j,j^\prime}
-{\cal W}^{-}_{j,j^\prime}-\left({\cal W}^{+}_{j,M}-{\cal W}^{-}_{j,M}\right)\right]\ .
\label{e17}
\end{eqnarray}
Once the non-equilibrium part, $g^\prime_j(t)$, of the total electron distribution function has
been solved from Eq.\,(\ref{e1}), the
transient drift velocity, $v_c(t)$, of the system can be calculated according to

\begin{equation}
\label{EQ:DRIFTVELOCITY}
v_{\rm c}(t)=\left[{\sum\limits_{j=1}^{N}\,f^{(0)}_j}\right]^{-1}\times
\left\{
\begin{array}{ll}
{\sum\limits_{j\neq M}\,\left(v_j -v_M\right)\,g^\prime_j(t)} & ,\ \ \ \  \mbox{semiconducting}\\
2\nu_F\sum\limits_{j=1}^{M}\,g^\prime_j(t) & ,\ \ \ \  \mbox{metallic}
\end{array}
\right.\ .
\end{equation}
It is noted that the thermal-equilibrium part of the electron distribution does not contribute to
the drift velocity.
The steady-state drift velocity $v_{\rm d}$ of electrons is given by
$v_{\rm c}(t)$ in the limit $t \to \infty$. Corresponding steady-state conduction
current is given by $I=en_{1D}v_{ \rm d}$. The differential electron mobility for
nonlinear transport is generalized to
$\mu_e=\partial v_{\rm d}/\partial\mathcal{F}_0$.
Numerical simulation of these quantities in specific ANRs is the subject of the next section.

\section{Numerical Results and Discussion}
\label{sec3}

\begin{figure}[p]
\epsfig{file=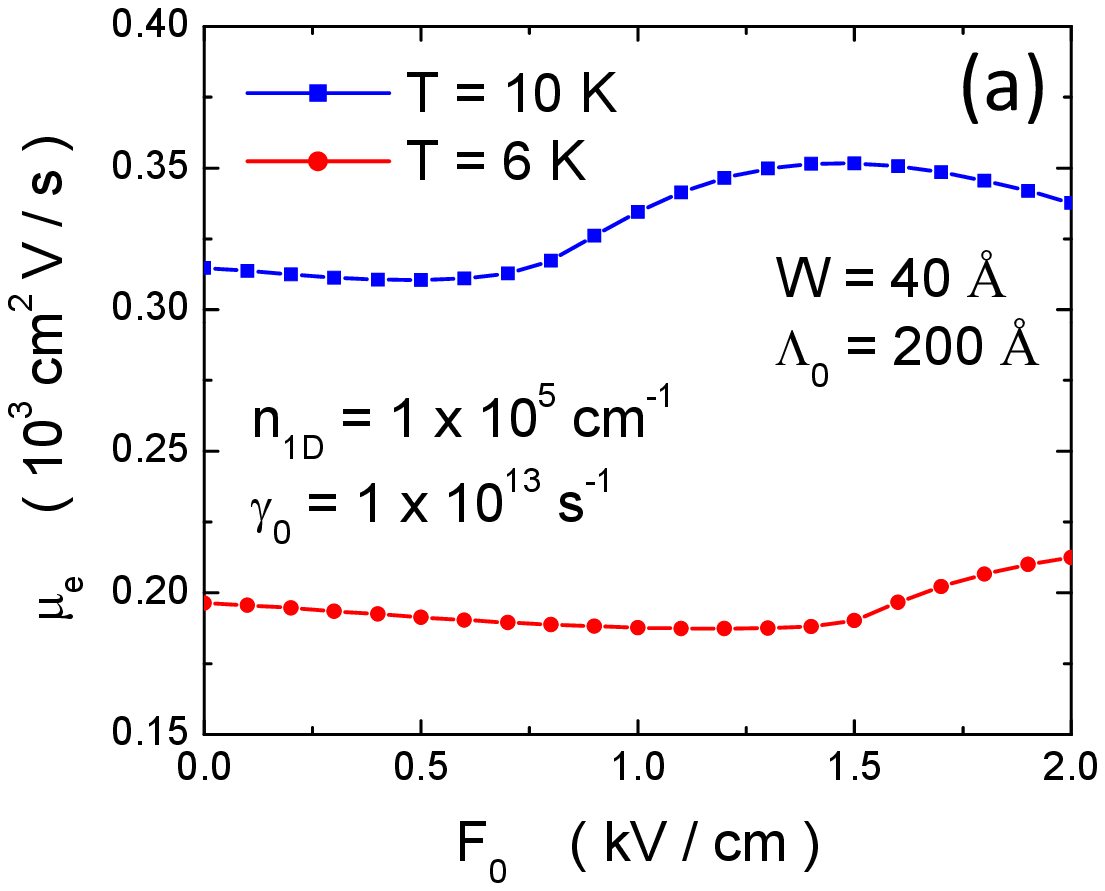,width=3.0in}
\epsfig{file=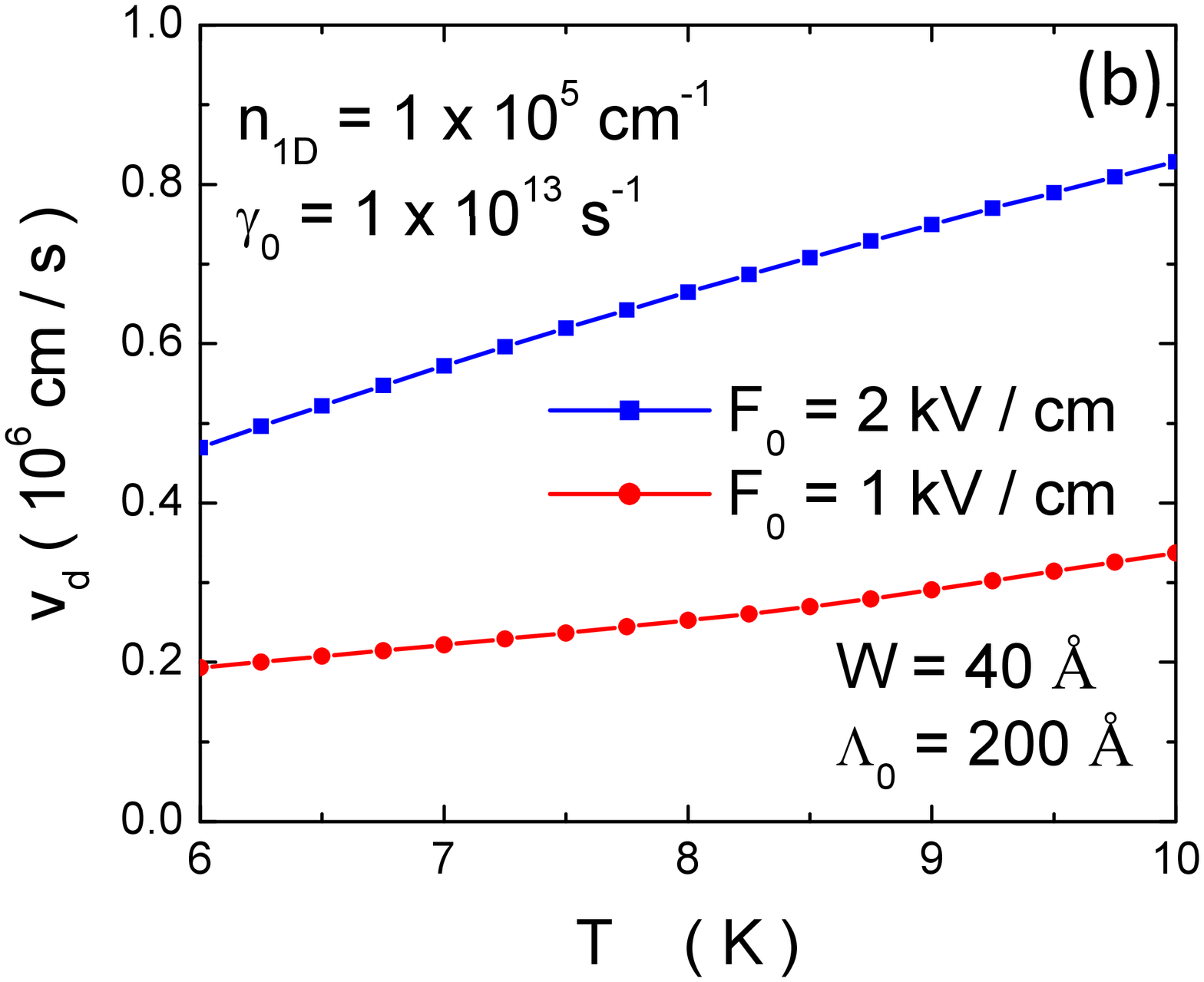,width=3.0in}\\
\vspace{0.5cm}
\epsfig{file=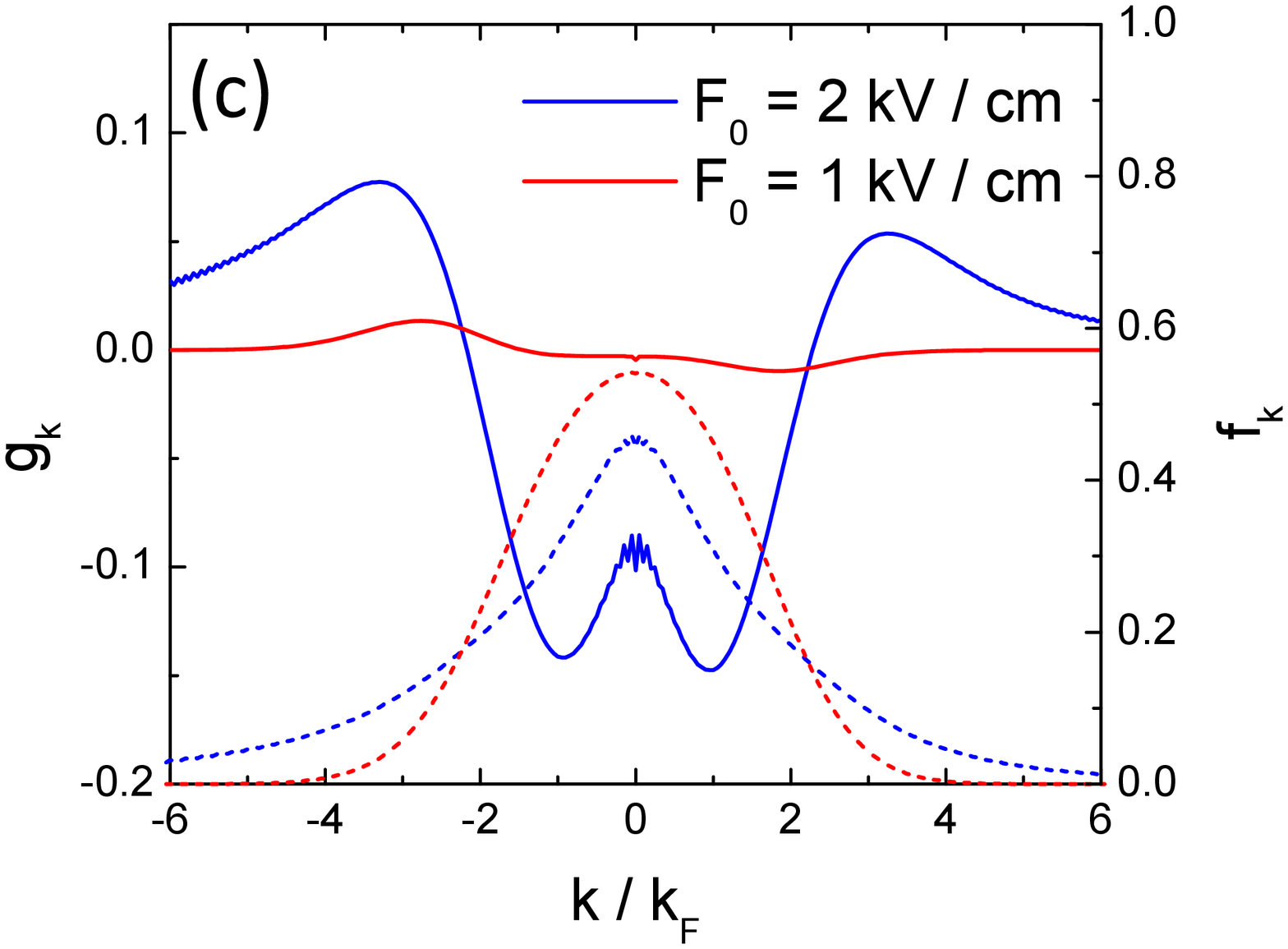,width=3.2in}
\epsfig{file=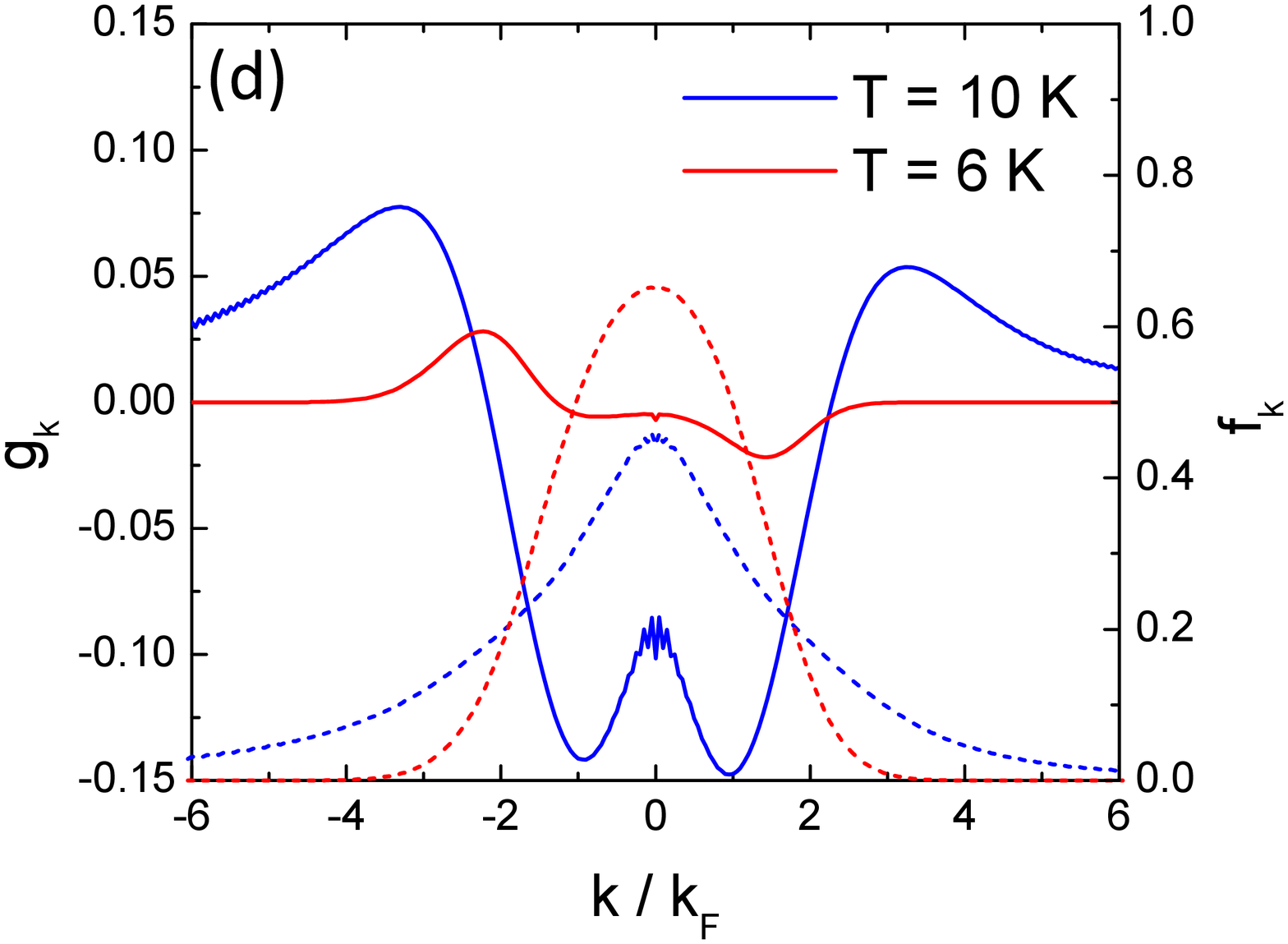,width=3.2in}
\caption{(Color online) (a) Calculated electron mobilities $\mu_{\rm e}$ as a function of applied
electric field ${\cal F}_0$ at $T=10$\,K (solid squares on blue curve) and $T=6$\,K
(solid circles on red curve); (b) Electron drift velocities $v_{\rm d}$ as a function of
temperature  $T$ at ${\cal F}_0=2$\,kV/cm (blue curve) and ${\cal F}_0=1$\,kV/cm (red curve);
(c) Non-equilibrium part of ($g_k$, left-hand scaled solid curves) and total ($f_k$, right-hand scaled
dashed curves) electron distribution functions at $T=10$\,K as functions of electron wave number
$k$ along the ribbon with ${\cal F}_0=2$\,kV/cm (blue curves) and ${\cal F}_0=1$\,kV/cm (red curves);
(d) $g_k$ (left-hand scaled solid curves) and $f_k$ (right-hand scaled dashed curves) with ${\cal F}_0=2$\,kV/cm
as functions of $k$ at $T=10$\,K (blue curves) and $T=6$\,K (red curves). The other parameters are
indicated directly in (a) and (b).}
\label{f1}
\end{figure}

Figure\ \ref{f1}(a) presents our  calculated electron mobilities $\mu_{\rm e}$ as
a function of applied electric field ${\cal F}_0$ at $T=10$\,K (blue curve)
and $T=6$\,K (red curve), respectively. Clearly, from Fig.\,\ref{f1}(a),  a strong
${\cal F}_0$-dependence for $\mu_{\rm e}$ appears at a lower value of ${\cal F}_0$
at higher temperature $T$ in the nanoribbon. This ${\cal F}_0$-dependent electron
mobility $\mu_{\rm e}$ has its physical origins  in the dynamical electron-phonon scattering
rate ${\cal W}_j^g(t)$ through its dependence on electron the distribution function
given in Eq.\,(\ref{e17}). Consequently, it is reasonable to
expect a lower threshold field, ${\cal F}^\ast$, for entering into a nonlinear transport regime (${\cal F}>{\cal F}^\ast$) due to
enhanced nonlinear phonon scattering at $T=10$\,K.
The value of ${\cal F}^\ast$ strongly depends on the parameters of the system, such as
$T$, $n_{\rm 1D}$, $\gamma_0$ and $\Lambda_0$, and
an analytic expression for ${\cal F}^\ast$ cannot be obtained in the nonlinear-transport regime.
On the other hand, as ${\cal F}_0\to 0$, $\mu_{\rm e}$
is larger at $T=10$\,K than  $T=6$\,K due to thermal population of
high-energy states with a large electron group velocity. The initial decrease of
$\mu_{\rm e}$ with ${\cal F}_0$ is attributed to the gradual increase of the frictional force from phonon
scattering by ${\cal F}_0$. At $T=10$\,K, $\mu_{\rm e}$ is roughly independent of
${\cal F}_0$ below $0.75$\,kV/cm (linear regime). However, $\mu_{\rm e}$ increases
significantly with ${\cal F}_0$ above $0.75$\,kV/cm (nonlinear regime). Eventually,
$\mu_{\rm e}$ decreases with ${\cal F}_0$ once it exceeds $1.5$\,kV/cm (heating regime),
leading to a saturation of the electron drift velocity. The electron group velocity
$|v_j|$ increases with the wave number for small $|k_j|$ values, as can be seen from
Eq.\,(\ref{EQ:DISPERSION}).
However, the increase of $|v_j|$ slows down toward its upper limit $\nu_F$ provided $|k_j|\gg\pi/3W$
but still within the single-subband regime.
The increase of $\mu_{\rm e}$ with ${\cal F}_0$ in the nonlinear regime comes from the
initially-heated electrons in high energy states with a larger group velocity,
while the successive decrease of $\mu_{\rm e}$ in the heating regime comes from the combination of the
upper limit $v_j\leq \nu_F$ and the dramatically increased phonon scattering. In Fig.\,\ref{f1}(b), the calculated electron
drift velocities $v_{\rm d}$ are plotted as a functions of temperature when
${\cal F}_0=2$\,kV/cm (blue curve) and ${\cal F}_0=1$\,kV/cm (red curve). The fact that
$\mu_{\rm e}$ increases with $T$ monotonically in both cases implies the electron scattering in
two samples is not dominated  by phonons but by impurities and line-edge roughness.
Different behaviors in the increase of $\mu_{\rm e}$ with temperature can be seen from Fig.\,\ref{f1}(b)
for linear and nonlinear electron transport. At ${\cal F}_0=2$\,kV/cm for the high-field nonlinear
transport, $v_{\rm d}$ (or $\mu_{\rm e}$) increases with $T$ sub-linearly. On the other hand,
$v_{\rm d}$ rises super-linearly with $T$ for the low-field linear transport at
${\cal F}_0=1$\,kV/cm. These different $T$ dependence in $\mu_{\rm e}$ for linear and nonlinear
transports can be directly related to the non-equilibrium part of the electron distribution
function $g_j$ presented in Figs.\,\ref{f1}(c) and \ref{f1}(d). The calculated $g_j$ (left scaled solid curves)
at $T=10$\,K, as well as the total electron distribution function $f_j$ (right scaled dashed curves),
are shown in Fig.\,\ref{f1}(c) as functions of electron wave number $k_j$ along the ribbon
when ${\cal F}_0=2$\,kV/cm (blue curves) and ${\cal F}_0=1$\,kV/cm (red curves).
The electron heating may be  visualized from Fig.\,\ref{f1}(c) when ${\cal F}_0=2$\,kV/cm
as a result of thermally driven electrons from low to high-energy states
by  heat generated from a frictional force\,\cite{heat} through nonlinear phonon scattering.
On the other hand, when ${\cal F}_0=1$\,kV/cm, electrons are only swept by elastic scattering
from the right Fermi edge to the left Fermi edge, corresponding to  linear transport.
Figure\ \ref{f1}(d) also allows a comparison between the calculated $g_j$ (left-hand scaled solid curves)
and $f_j$ (right scaled dashed curves) with ${\cal F}_0=2$\,kV/cm as functions of $k_j$ at
$T=10$\,K (blue curves) and $T=6$\,K (red curves). One may conclude from Fig.\,\ref{f1}(d)
that the nonlinear phonon scattering becomes important at $T=10$\,K, while the elastic scattering
of electrons dominates at $T=6$\,K, similar to the explanation given for Fig.\,\ref{f1}(c).
\medskip

\begin{figure}[p]
\begin{center}
\epsfig{file=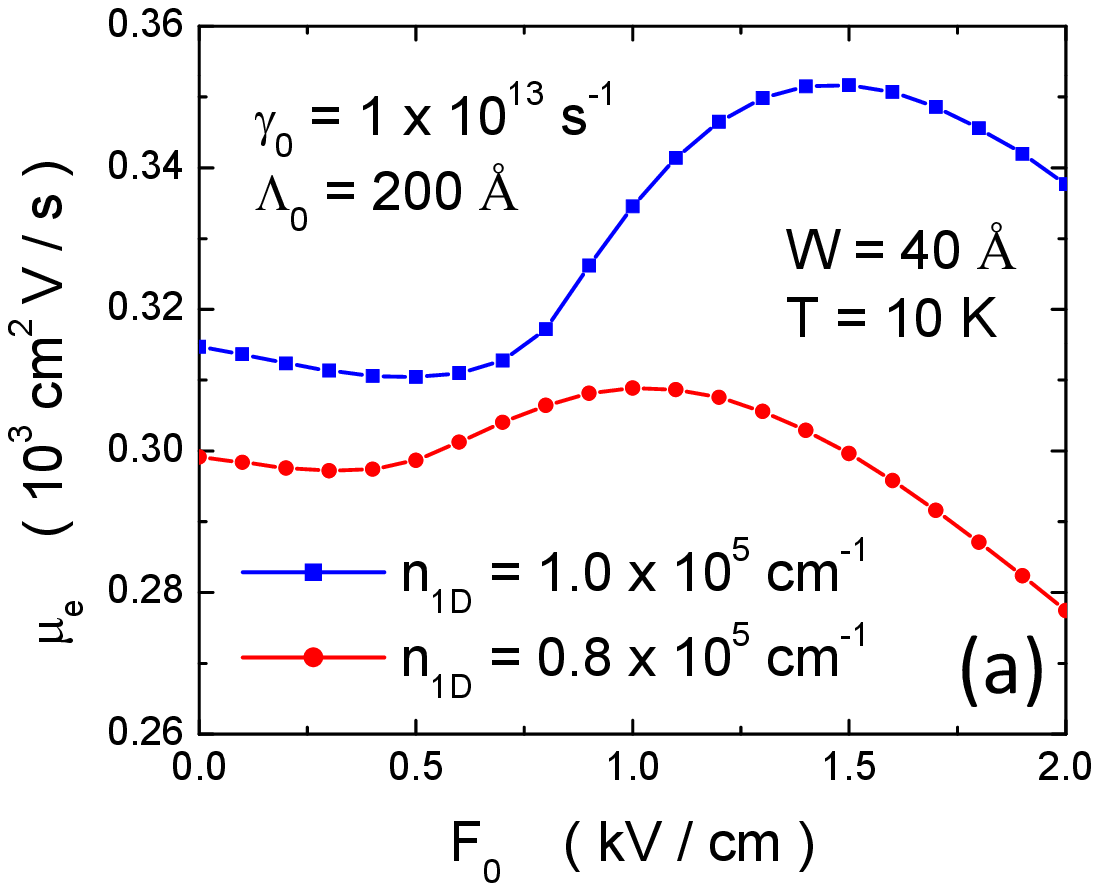,width=3.0in}
\epsfig{file=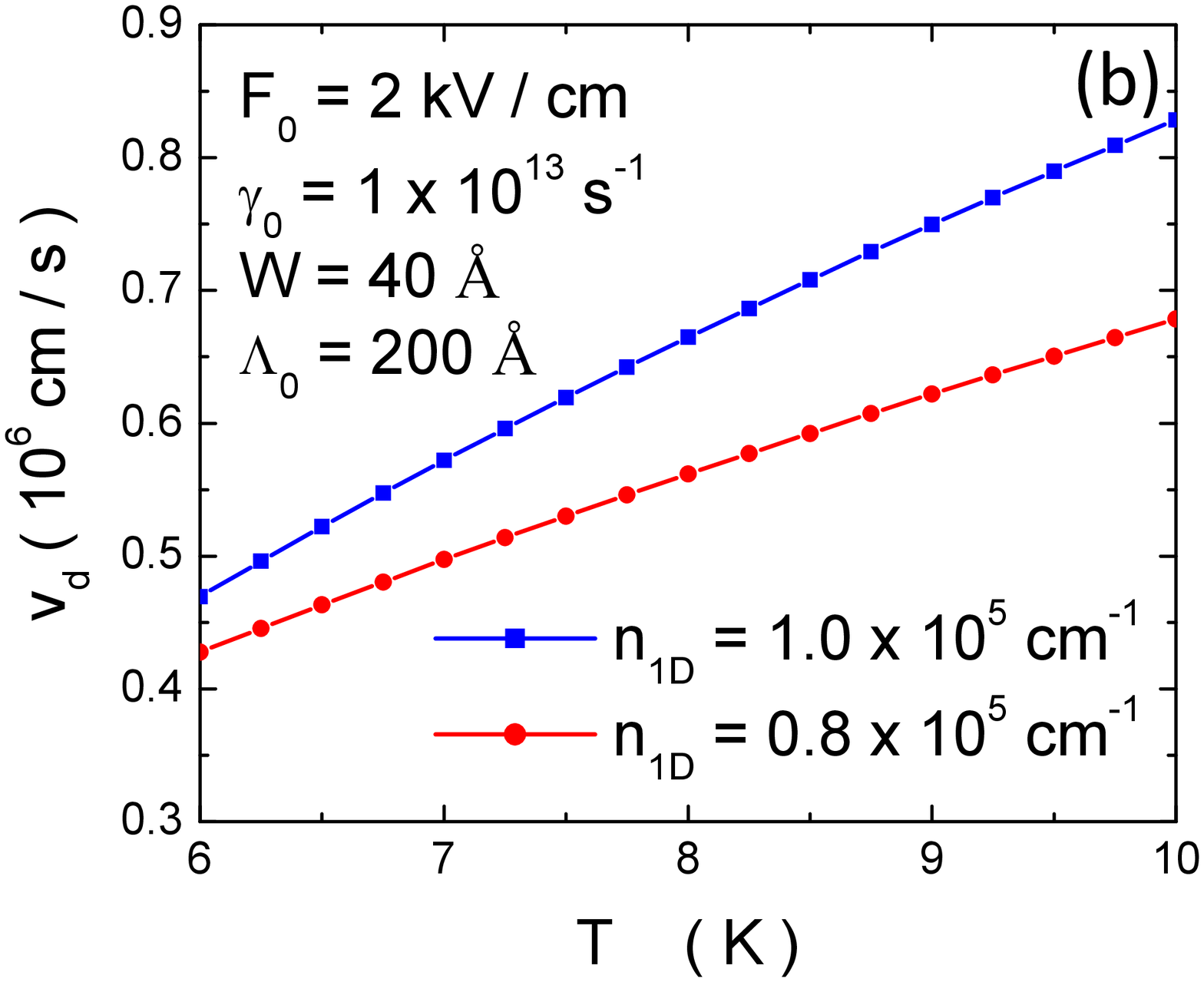,width=3.0in}\\
\vspace{0.5cm}
\epsfig{file=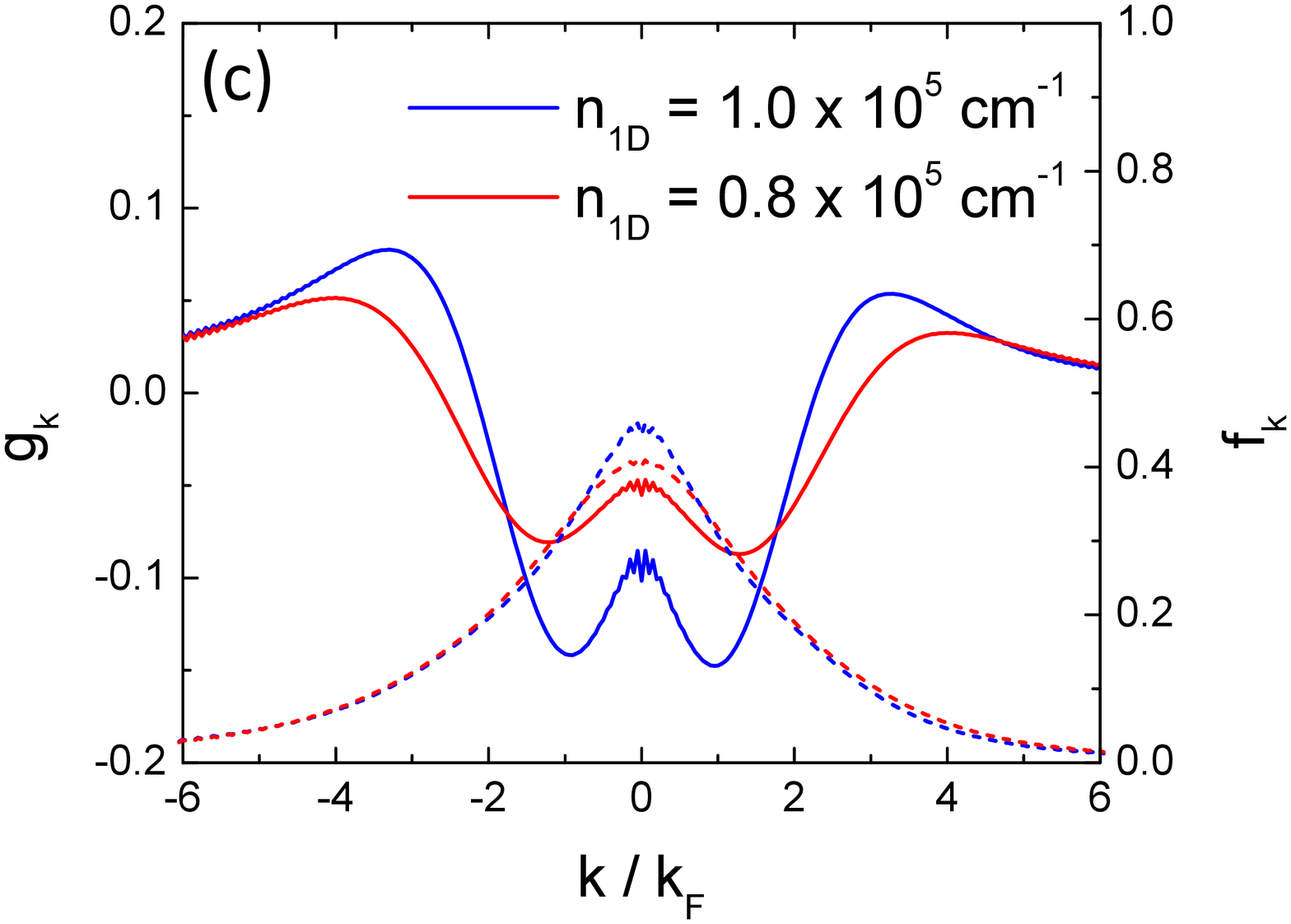,width=3.2in}
\end{center}
\caption{(Color online) (a) Calculated $\mu_{\rm e}$ as a function of ${\cal F}_0$ at $T=10$\,K with
$n_{\rm 1D}=1.0\times 10^{5}$\,cm$^{-1}$ (solid squares on blue curve)
and $n_{\rm 1D}=0.8\times 10^{5}$\,cm$^{-1}$ (solid circles on red curve);
(b) $v_{\rm d}$ as a function of $T$ with ${\cal F}_0=2$\,kV/cm for
$n_{\rm 1D}=1.0\times 10^{5}$\,cm$^{-1}$ (solid squares on blue curve)
and $n_{\rm 1D}=0.8\times 10^{5}$\,cm$^{-1}$ (solid circles on red curve);
(c) $g_k$ (left-hand scaled solid curves) and $f_k$ (right-hand scaled dashed curves) as a function
of $k$. Here, the cases with $n_{\rm 1D}=1.0\times 10^5$\,cm$^{-1}$ and
$n_{\rm 1D}=0.8\times 10^5$\,cm$^{-1}$ are represented by blue and red curves, respectively.
The other parameters are indicated in (a) and (b).}
\label{f2}
\end{figure}

We present, in Fig.\,\ref{f2}(a), our numerical results for $\mu_{\rm e}$ as a
function of ${\cal F}_0$ at $T=10$\,K for different electron densities
$n_{1D}=1.0\times 10^5$\,cm$^{-1}$ (blue curve) and $n_{1D}=0.8\times 10^5$\,cm$^{-1}$
(red curve). As ${\cal F}_0\to 0$, $\mu_{\rm e}$ is slightly increased for a higher value of
$n_{1D}$ due to introducing high-energy occupied states with a large group velocity.
It is further found from Fig.\,\ref{f2}(a) that a low electron density produces a small
${\cal F}^\ast$ in the system. This is due to the fact that a small minimum energy
$\hbar\omega_q\approx 2\hbar c_sk_{\rm F}$ is required for initializing phonon scattering
in a low density sample, which enhances the nonlinear phonon scattering at a chosen temperature.
However, the enhancement of $\mu_{\rm e}$ due to ${\cal F}_0$ is large in a high
density sample due to increased initial electron heating. In Fig.\,\ref{f2}(b), $v_{\rm d}$
is presented at ${\cal F}_0=2$\,kV/cm as a function of $T$ for $n_{1D}=1.0\times 10^5$\,cm$^{-1}$
(blue curve) and $n_{\rm 1D}=0.8\times 10^5$\,cm$^{-1}$ (red curve). Although we still see
$\mu_{\rm e}$ increasing sub-linearly with $T$, as explained in Fig.\,\ref{f1}(b), the rate of
$\mu_{\rm e}$ enhancement is large for a high-$n_{1D}$ sample when $T$ is not low, where the electron
heating is significant. This observation is complemented by the results shown in Fig.\,\ref{f2}(c),
where electron heating becomes more significant for a high density sample although the nonlinear
phonon scattering is seen to be important in both low and high density samples.
This enhancement of $\mu_{\rm e}$ with $T$ at high electron densities
is due to more low-energy electrons being available for thermal excitation to occupy
high-energy states through electron heating.
\medskip

\begin{figure}[p]
\begin{center}
\epsfig{file=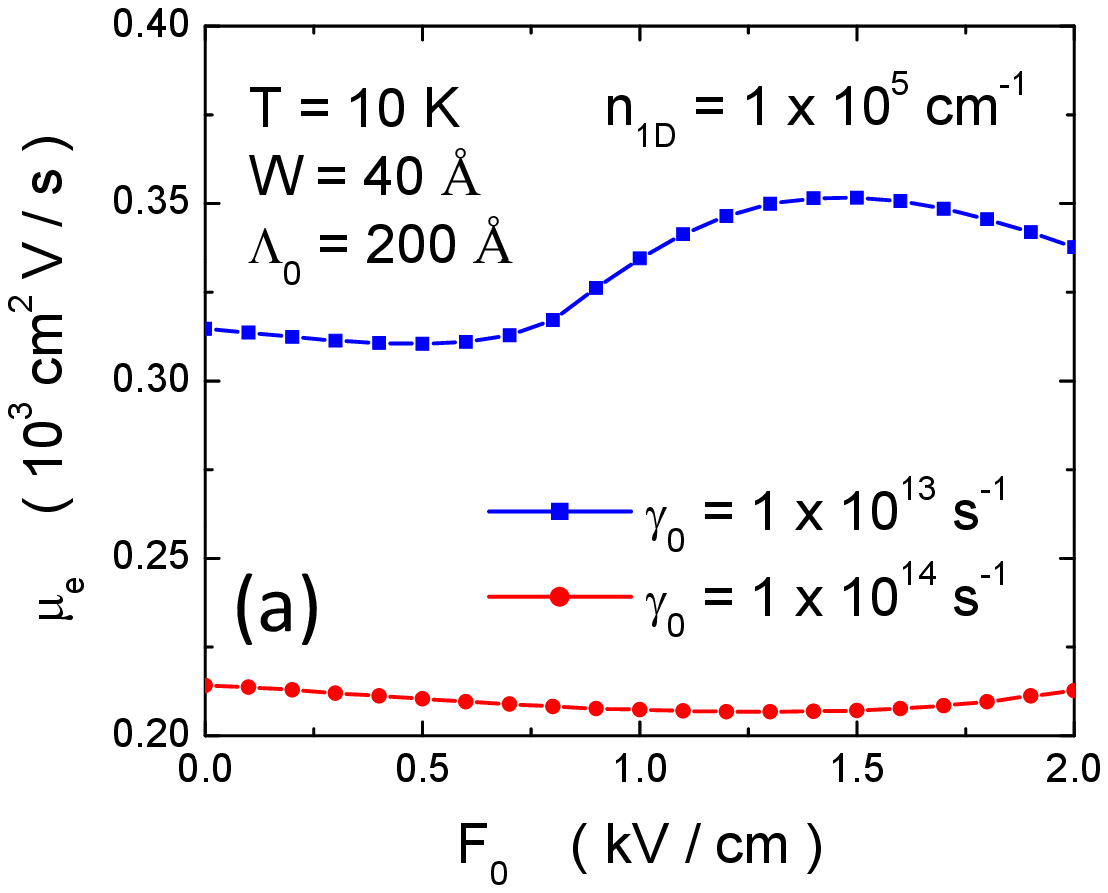,width=3.0in}
\epsfig{file=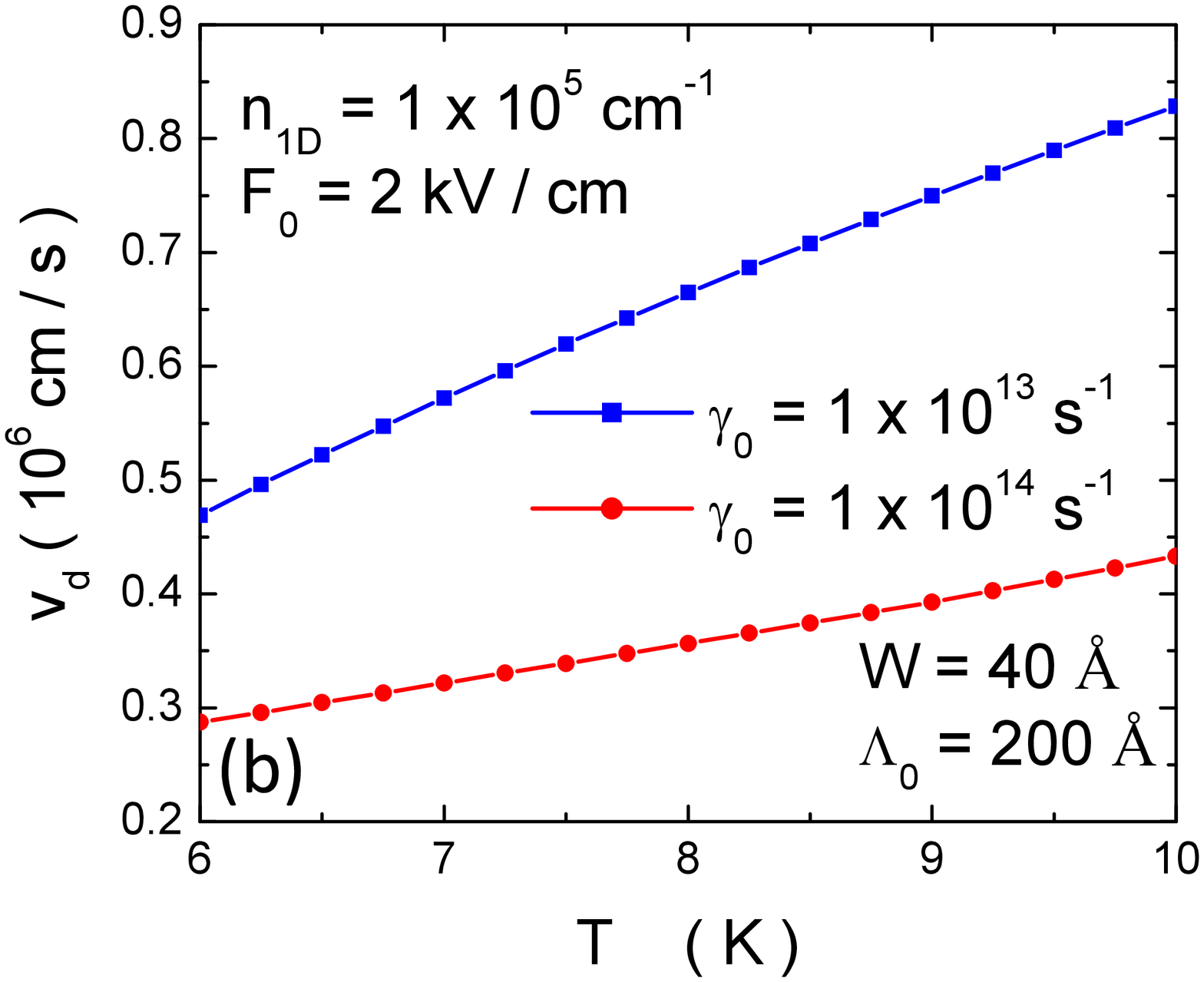,width=3.0in}\\
\vspace{0.5cm}
\epsfig{file=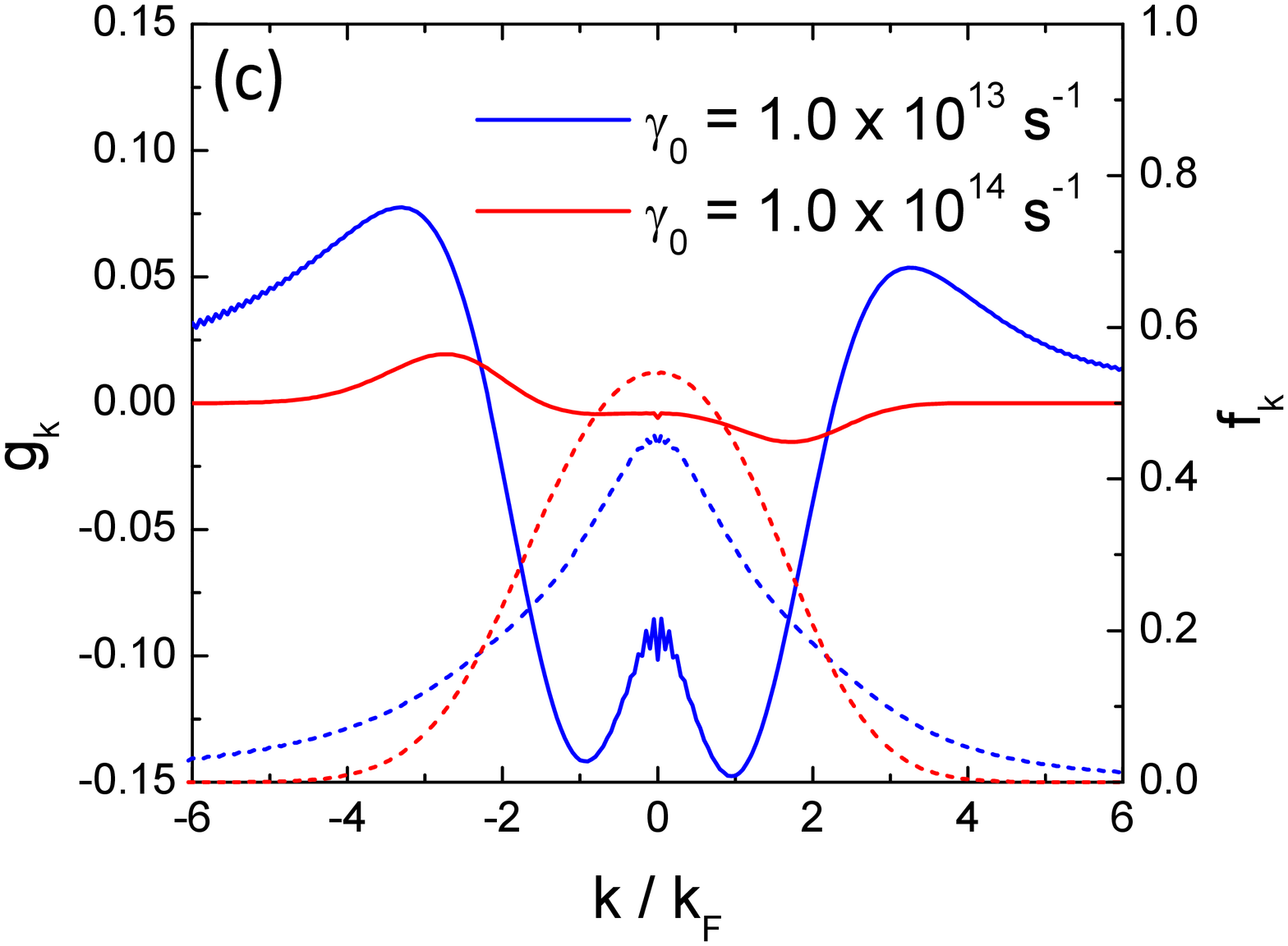,width=3.2in}
\end{center}
\caption{(Color online) (a) $\mu_{\rm e}$ as a function of ${\cal F}_0$ at $T=10$\,K with
$\gamma_0=1.0\times 10^{13}$\,s$^{-1}$ (solid squares on blue curve)
and $\gamma_0=1.0\times 10^{14}$\,s$^{-1}$ (solid circles on red curve);
(b) $v_{\rm d}$ as a function of $T$ with ${\cal F}_0=2$\,kV/cm for
$\gamma_0=1.0\times 10^{13}$\,s$^{-1}$ (solid squares on blue curve)
and $\gamma_0=1.0\times 10^{14}$\,s$^{-1}$ (solid circles on red curve);
(c) $g_k$ (left-hand scaled solid curves) and $f_k$ (right-hand scaled dashed curves) as a function of $k$.
Here, the cases with $\gamma_0=1.0\times 10^{13}$\,s$^{-1}$ and $\gamma_0=1.0\times 10^{14}$\,s$^{-1}$
are represented by blue and red curves, respectively.
The other parameters are indicated in (a) and (b).}
\label{f3}
\end{figure}

The effects due to impurity scattering are compared in Figs.\,\ref{f3}(a)-(c).
In Fig.\,\ref{f3}(a), we present a comparison of mobilities as a function of
${\cal F}_0$ at $T=10$\,K for $\gamma_0=1.0\times 10^{13}$\,s$^{-1}$ (blue curve) and
$\gamma_0=1.0\times 10^{14}$\,s$^{-1}$ (red curve). As ${\cal F}_0\to 0$, $\mu_{\rm e}$ is
greatly reduced by strong impurity scattering with $\gamma_0=1.0\times 10^{14}$\,s$^{-1}$ in
the linear regime. In addition, for $\gamma_0=1.0\times 10^{14}$\,s$^{-1}$, ${\cal F}^\ast$
is pushed upward from about $0.75$\,kV/cm to a  higher value at $1.75$\,kV/cm, leaving us
with a roughly ${\cal F}_0$-independent $\mu_{\rm e}$ in this case for the whole field range
shown in this figure. The comparison for $T$-dependence of $\mu_{\rm e}$ is presented
in Fig.\,\ref{f3}(b) for ${\cal F}_0=2$\,kV/cm, where a sub-linear increase of $\mu_{\rm e}$ with
$T$ for weak impurity scattering is switched to a super-linear relation in the strong impurity
scattering case. The effect of impurity scattering can also be seen from the calculated $g_k$ and
$f_k$ as functions of $k$ at $T=10$\,K and ${\cal F}_0=2$\,kV/cm in Fig.\,\ref{f3}(c), where the
electron heating with $\gamma_0=1.0\times 10^{13}$\,s$^{-1}$ through nonlinear phonon scattering
is completely masked by the strong impurity scattering at $\gamma_0=1.0\times 10^{14}$\,s$^{-1}$, leading
to relative  cooling of conduction electrons in a nanoribbon in the linear regime.
\medskip

At relatively low temperatures $T\le 10$\,K and low concentration of electrons ($n_{\rm 1D}=10^5$\,cm$^{-1}$) in the lowest conduction band, as well
as medium impurity concentration $n_{\rm 2D}=10^{10}$\,cm$^{-2}$, the single conduction band approximation can be justified for graphene nanoribbons.
Indeed, the energy separation between the first and second conduction subband is $\pi\hbar\nu_F/3W\approx 0.136$\,eV for $W=50$\,\AA.
On the other hand, elastic and inelastic decay rate $\hbar \gamma_{1}<\hbar \gamma_{0}=0.065$\,eV, as well as the Fermi energy
$\varepsilon_F(\pi n_{1D}/2)\approx 0.01$\,eV, are much smaller.\,\footnote{$\hbar\gamma_{1}<\hbar\gamma_{0}$ is valid for $\Lambda_0/W <6$}
More generally speaking, the model is valid provided that the scattering parameters
and Fermi energy are of the order of the subband gap $E_g=1.38/W$\,eV with $W$ expressed in nanometers.
\medskip

\begin{figure}[p]
\begin{center}
\epsfig{file=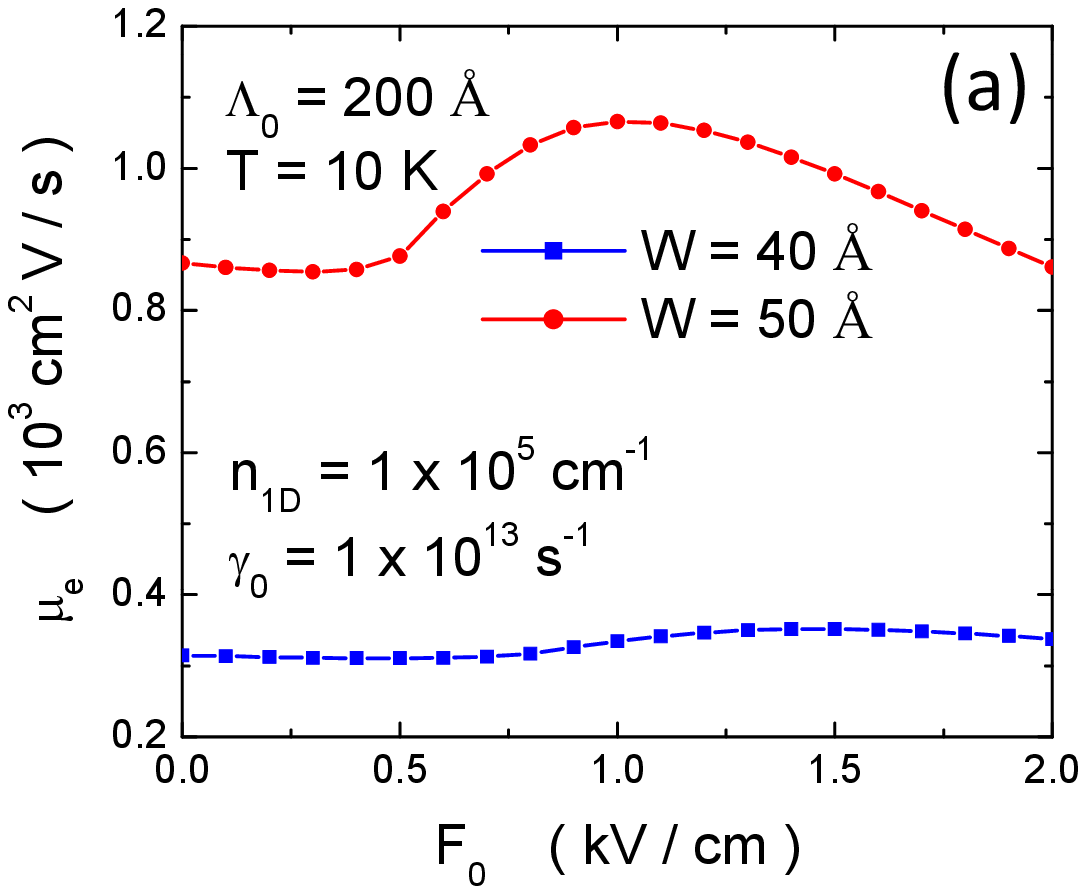,width=3.0in}
\epsfig{file=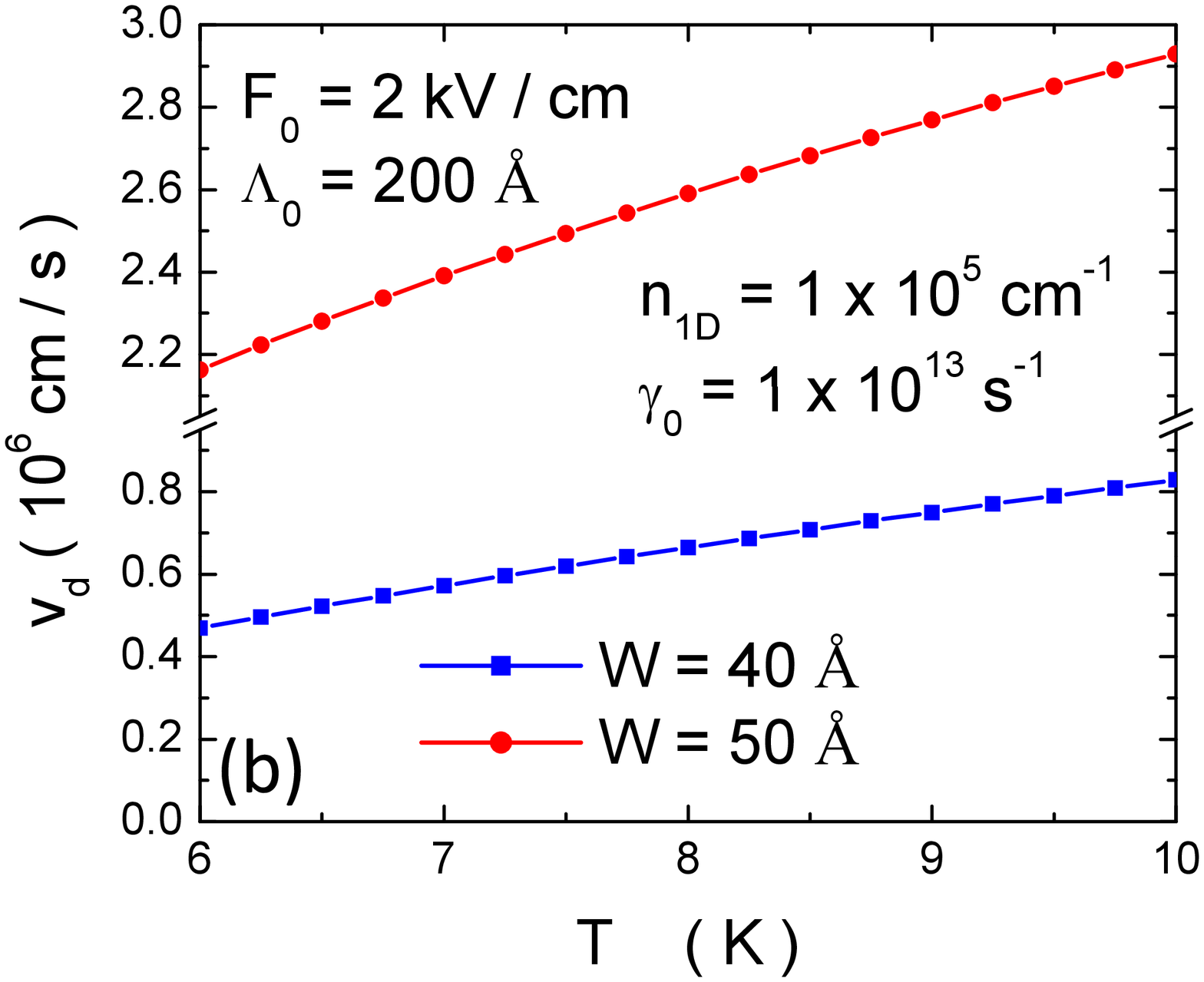,width=3.0in}\\
\vspace{0.5cm}
\epsfig{file=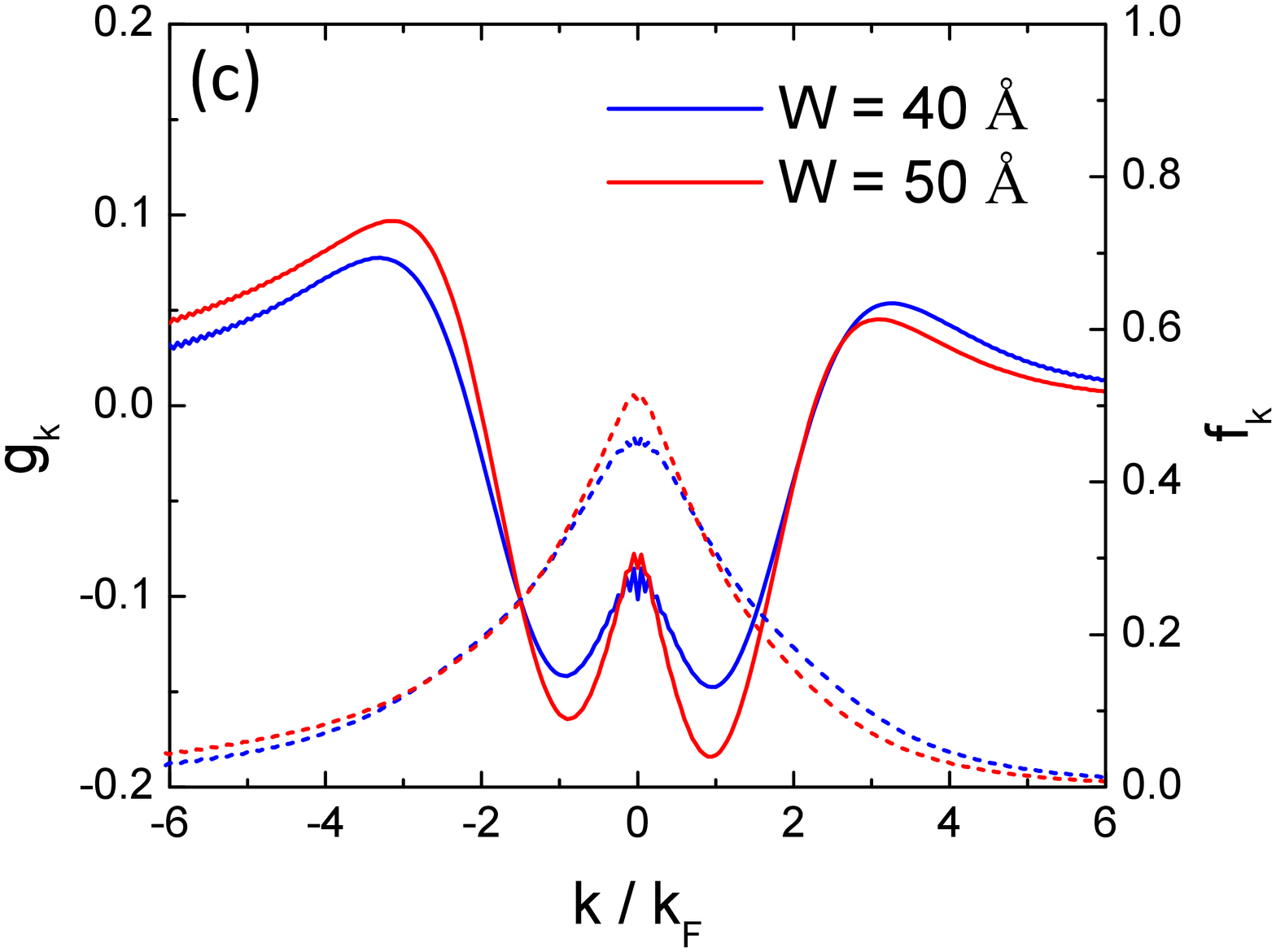,width=3.2in}
\end{center}
\caption{(Color online) (a) $\mu_{\rm e}$ as a function of ${\cal F}_0$ at $T=10$\,K with
$W=40$\,\AA\ (solid squares on blue curve) and $W=50$\,\AA\ (solid circles on red curve);
(b) $v_{\rm d}$ as a function of $T$ with ${\cal F}_0=2$\,kV/cm for $W=40$\,\AA\ (solid squares
on blue curve) and $W=50$\,\AA\ (solid circles on red curve);
(c) $g_k$ (left-hand scaled solid curves) and $f_k$ (right-hand  scaled dashed curves) as a function of $k$.
Here, the cases with $W=40$\,\AA\ and $W=50$\,\AA\ are represented by blue and red curves, respectively.
The other parameters are indicated in (a) and (b).}
\label{f4}
\end{figure}

The increase of ribbon's width $W$ will enlarge the electron group velocity $v_j<v_M\sim[1+(2/3 n_{1D} W)^2]^{-1/2}$,
as shown by Eq.\,(\ref{EQ:DISPERSION}), and significantly decrease
the effect of line-edge roughness scattering at the same time, as can be seen from the factor
$\gamma_1\sim(\pi\nu_F/3W)^2\,(\delta b/W)^2$ in Eq.\,(\ref{EQ:elasticrate}).
Fixing $\Lambda_0 = 200$\,\AA\ and making ${\cal F}_0\to 0$, we do see from Fig.\,\ref{f4}(a) that $\mu_{\rm e}$ is dramatically
increased when $W$ is changed
from $40$\,\AA\ ($\gamma_{1}=1.7\times 10^{13}$\,s$^{-1}$) to $50$\,\AA\ ($\gamma_{1}=4.28\times 10^{13}$\,s$^{-1}$)
due to enlarged group velocity of electrons and reduced line-edge roughness scattering.
In addition, the magnitude in the enhancement of $\mu_{\rm e}$ with ${\cal F}_0$ is
large in the sample with $W=50$\,\AA\ because the weakened elastic scattering, and therefore,
the relatively strong initial heating of electrons by phonon scattering to high energy states with a larger group velocity.
This is also reflected in the temperature dependence of $v_{\rm d}$ presented in Fig.\,\ref{f4}(b).
However, the rate for $\mu_{\rm e}$ enhancement with $T$ is almost the same in the two cases,
implying a similar magnitude of nonlinear phonon scattering of electrons in both samples.
This can be very well justified by the calculated $g_j$ shown in Fig.\,\ref{f4}(c),
where the increase of $W$ from $40$\,\AA\ to $50$\,\AA\ only slightly enhances the
asymmetry of $g_j$ with respect to $\pm k_j$, but the unique feature of electron heating is largely unchanged.
\medskip

\begin{figure}[p]
\begin{center}
\epsfig{file=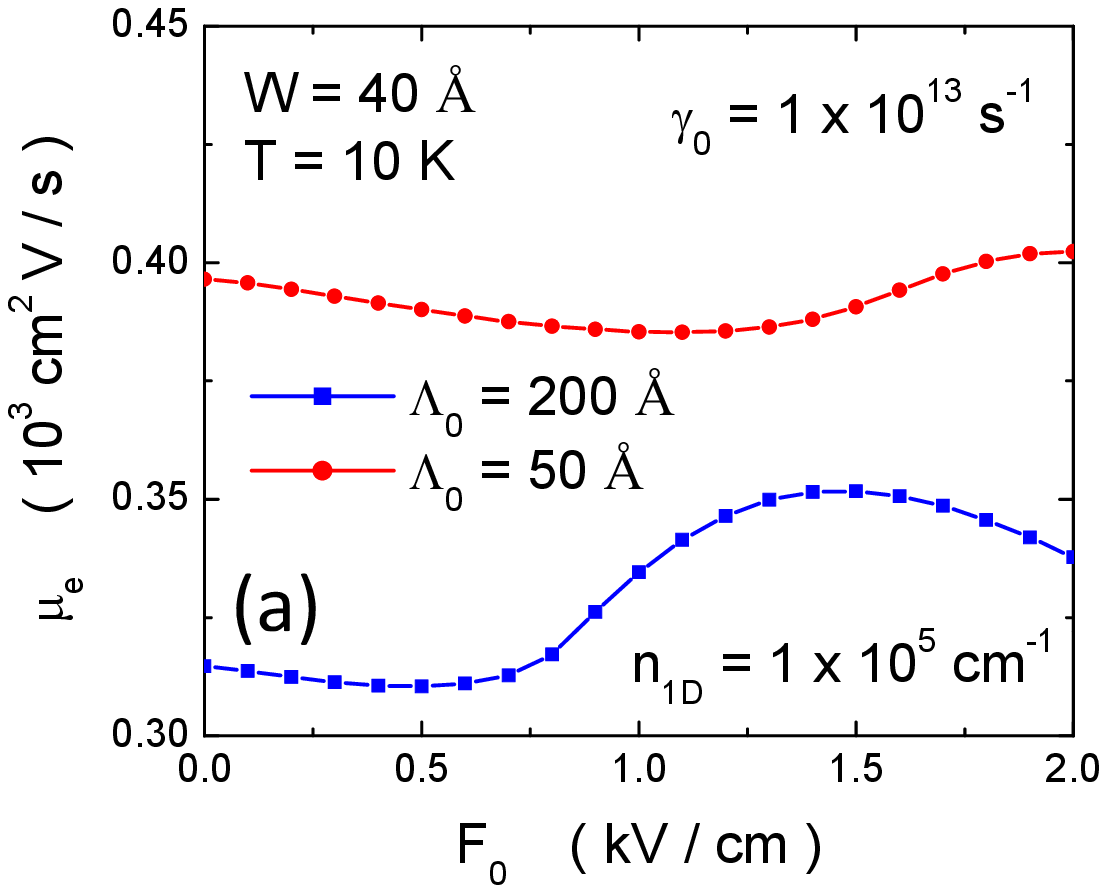,width=3.0in}
\epsfig{file=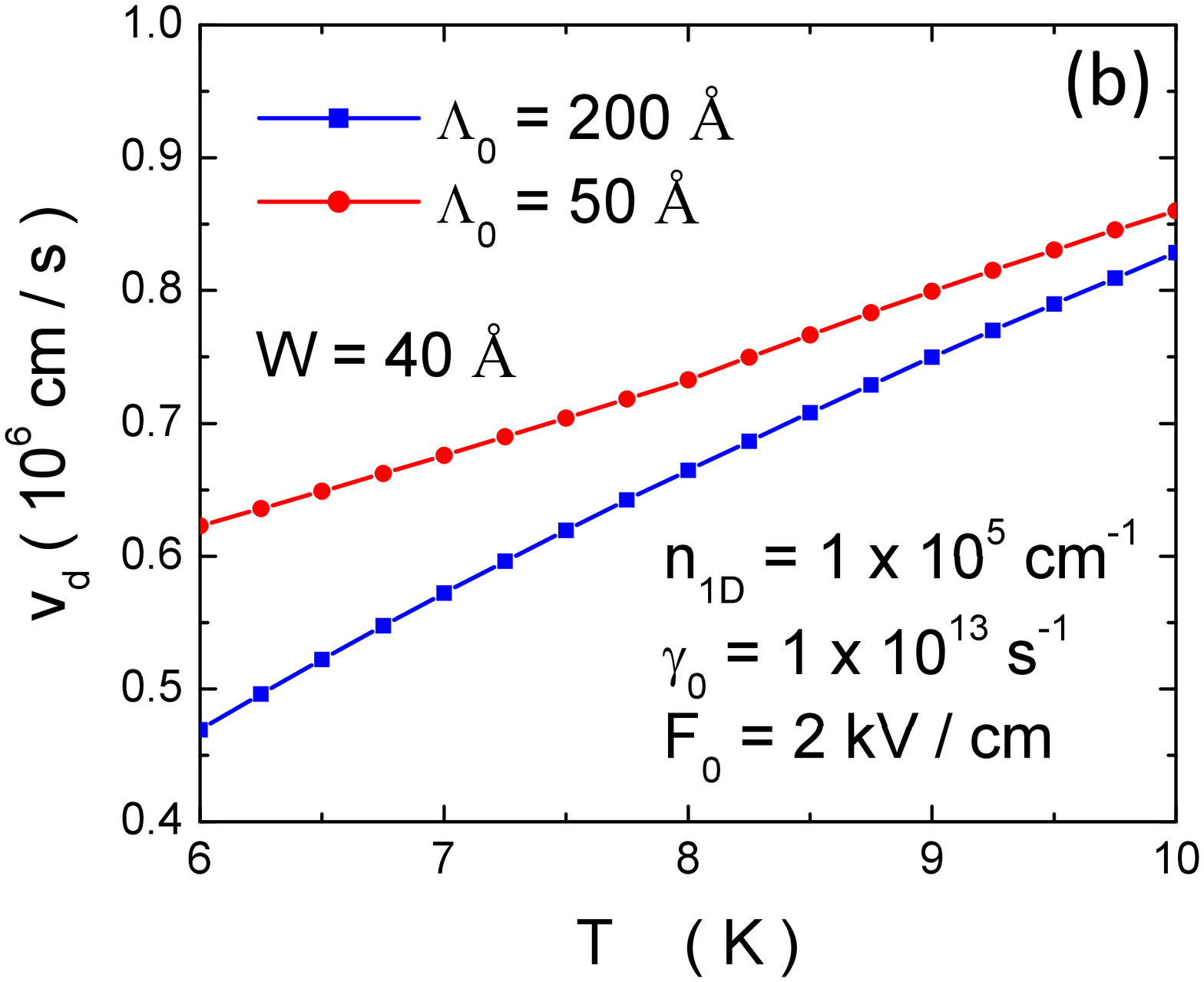,width=3.0in}\\
\vspace{0.5cm}
\epsfig{file=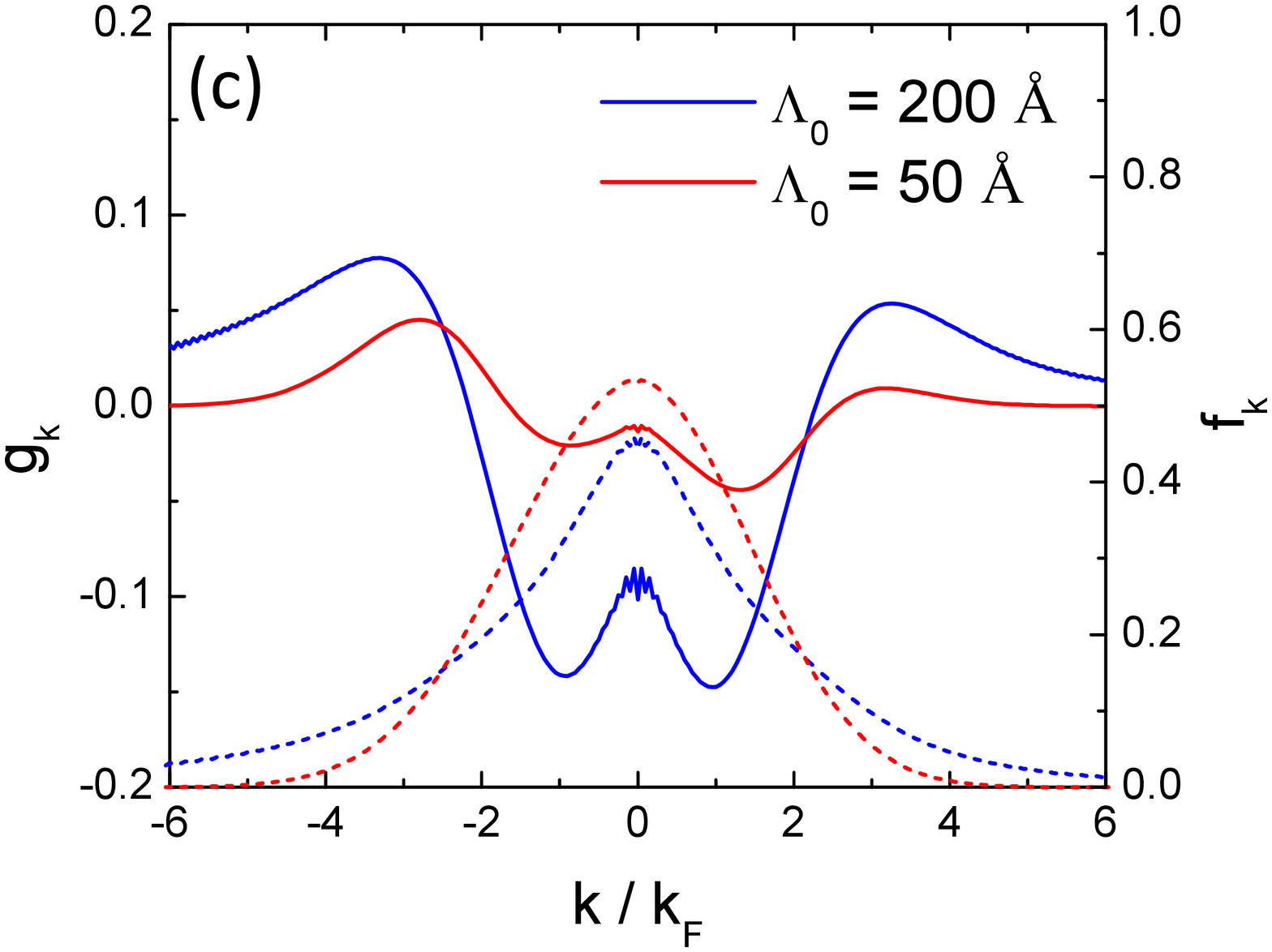,width=3.2in}
\end{center}
\caption{(Color online) (a) $\mu_{\rm e}$ as a function of ${\cal F}_0$ at $T=10$\,K with
$\Lambda_0=200$\,\AA\ (solid squares on blue curve) and $\Lambda_0=50$\,\AA\ (solid circles on
red curve); (b) $v_{\rm d}$ as a function of $T$ with ${\cal F}_0=2$\,kV/cm for
$\Lambda_0=200$\,\AA\ (solid squares on blue curve) and $\Lambda_0=50$\,\AA\ (solid circles on red curve);
(c) $g_k$ (left-hand scaled solid curves) and $f_k$ (right-hand scaled dashed curves) as a function of $k$.
Here, the cases with $\Lambda_0=200$\,\AA\ and $\Lambda_0=50$\,\AA\ are represented by blue and red curves, respectively.
The other parameters are indicated in (a) and (b).}
\label{f5}
\end{figure}

Finally, the effect of correlation length for the line-edge roughness on the transport is
demonstrated in Figs.\,\ref{f5}(a)-(c) by fixing $W=50$\,\AA\ and changing $\Lambda_0$ from $200$\,\AA\ to $50$\,\AA.
As shown in Eq.\,(\ref{EQ:elasticrate}), the line-edge roughness scattering can be either reduced or
enhanced by decreasing $\Lambda_0$, depending on $|k_j|\ll 1/2\Lambda_0$ or $|k_j|\gg 1/2\Lambda_0$.
For our chosen sample with $n_{\rm 1D}=1.0\times 10^5$\,cm$^{-1}$, we find that the condition
$|k_j|\ll 1/2\Lambda_0$ is satisfied in the low-field limit ($|k_j|\sim k_{\rm F}$), while
$|k_j|\gg 1/2\Lambda_0$  holds for the high field limit due to electron heating.
Therefore, we find from Fig.\,\ref{f5}(a) that $\mu_{\rm e}$ is increased as ${\cal F}_0\to 0$ when $\Lambda_0$
is reduced to $50$\,\AA\ in the low field regime. However, the value of ${\cal F}^\ast$ for $\mu_{\rm e}$
is pushed upward for $\Lambda_0=50$\,\AA\ in the high-field regime, which is further accompanied by a
reduced magnitude in the enhancement of $\mu_{\rm e}$ with ${\cal F}_0$.
This anomalous feature associated with reducing $\Lambda_0$ also has a profound
impact on the $T$-dependence of $\mu_{\rm e}$ as shown in Fig.\,\ref{f5}(b),
where the increasing rate of $\mu_{\rm e}$ with $T$ in the high field regime
becomes much lower with $\Lambda_0=50$\,\AA\ than for $\Lambda_0=200$\,\AA. In
addition, the calculated $g_j$ in Fig.\,\ref{f5}(c) displays an anomalous  cooling
behavior (with a smaller spreading of $f_j$ in the $k_j$ space) in the high field regime
for conduction electrons as $\Lambda_0$ is reduced to $50$\,\AA, similar to the results
shown in Fig.\,\ref{f3}(c).
\medskip

We have performed similar calculations with metallic \footnote{Two carbon atoms have been added/subtracted to the semiconducting ribbon across its width.}
ribbons of comparable width $W=50\pm5$\,\AA\ and found
no perceptible difference in the nonlinear electron transport and heating provided that the rest of the parameters
(electron and impurities concentrations and decay rates) are kept the same.

\section{Concluding Remarks}
\label{sec4}

In conclusion, we have found there is a minimum electron mobility just before a threshold for
an applied electric field when entering into the nonlinear transport regime, which
is attributed to the gradual build-up of a frictional force from phonon scattering by the applied field. We
also predict a field-induced mobility enhancement right after this threshold value,
which is regarded as a consequence of initially-heated electrons in high energy states with a larger group velocity in an elastic-scattering
dominated graphene nanoribbon. Moreover, we have discovered that this mobility enhancement
reaches a maximum in the nonlinear transport regime as a combined result of an upper limit
for the group velocity in a nanoribbon and a field-induced dramatically-increased phonon scattering in the system.
Finally, the threshold field can be pushed upward and the magnitude in the mobility enhancement
can be reduced simultaneously by a small correlation length for the line-edge roughness in
the high-field limit due to the occupation of high-energy states by field-induced electron heating.
\medskip

When the electron density is increased in a graphene nanoribbon, multi-subband transport occurs, the field-induced mobility enhancement is expected to be reduced,
and the effect of electron-electron scattering needs to be included. When the lattice
temperature becomes high, on the other hand, both the optical phonon and inter-valley
scattering should be considered. As the width of a nanoribbon is modulated, a periodic
potential along the ribbon will form, leading to a graphene nanoribbon superlattice with
additional miniband gap opening at the Brillouin zone boundaries. The results of the current
research is expected to be very useful for the understanding and  design of high-power and high-speed
graphene nanoribbon emitters and detectors in the terahertz frequency range.

\begin{acknowledgments}
This research was supported by contract \# FA 9453-07-C-0207 of AFRL. DH would like
to thank the Air Force Office of Scientific Research
(AFOSR) for its support.
\end{acknowledgments}

\begin{references}
\bibitem{art10}A. K. Geim and K. S. Novosolev, Nat. Mater. {\bf 6}, 183 (2007).

\bibitem{art11}K. S. Novosolev, A. K. Geim, S. V. Morozov, D. Jiang, M. I. Katsnelson,
I. V. Grigorieva, S. V. Dubonos, and A. A. Firsov, Nat. (London) {\bf 438}, 197 (2005).

\bibitem{art12}C. Berger, Z. Song, X. Li, X. Wu, N. Brown, C. Naud, D. Mayou, T. Li,
J. Hass, A. N. Marchenkov, E. H. Conrad, P. N. First, and W. A. de Heer, Sci. {\bf 312}, 1191 (2006).

\bibitem{neto}A. H. C. Neto, F. Guinea, N. M. R. Peres, K. S. Nonoselov, and A. K. Geim,
\rmp {\bf 81}, 109 (2009).

\bibitem{art13}P. Avouris, Z. Chen, and V. Perebeinos, Nat. Nanotechnol. {\bf 2}, 605 (2007).

\bibitem{art1}T. Ando, J. Phys. Soc. Jpn. {\bf 75}, 074716 (2006).

\bibitem{art2}J. H. Chen, C. Jang, S. Adam, M. S. Fuhrer, E. D. Williams, and M. Ishigami,
Nat. Phys. {\bf 4}, 377 (2008).

\bibitem{art3}H. M. Dong, W. Xu, Z. Zeng, T. C. Lu, and F. M. Peeters, \prb {\bf 77}, 235402 (2008).

\bibitem{art4}N. M. R. Peres, J. M. B. Lopes dos Santos, and T. Stauber, \prb
{\bf 76}, 073412 (2007); also see T. Stauber, N. M. R. Peres, and F. Guinea,
{\em ibid.} {\bf 76}, 205423 (2007).

\bibitem{art5}V. V. Cheianov and V. I. Fal\'ko, \prl {\bf 97}, 226801 (2006).

\bibitem{art7}W. Xu, F. M. Peeters, and T. C. Lu, \prb {\bf 79}, 073403 (2009).

\bibitem{art8}S. Y. Liu, X. L. Lei, and N. J. M. Horing, J. Appl. Phys. {\bf 104}, 043705 (2008).

\bibitem{art14}Y.-M. Lin, C. Dimitrakopoulos, K. A. Jenkins, D. B. Farmer, H.-Y. Chiu,
A. Grill, and P. Avouris, Sci. {\bf 327}, 662 (2010).

\bibitem{art15}T. Mueller, F. Xia, and P. Avouris, Nat. Photonics {\bf 4}, 297 (2010).

\bibitem{art6}Y.-M. Lin, V. Perebeinos, Z. Chen, and P. Avouris, \prb {\bf 78} 161409 (2008).

\bibitem{fang}T. Fang, A. Konar, H. Xing, and D. Jena, \prb {\bf 78}, 205403 (2008).

\bibitem{art9}X. Wang, Y. Ouyang, X. Li, H. Wang, J. Guo, and H. Dai, \prl {\bf 100}, 206803 (2008).

\bibitem{huang1}D. H. Huang and G. Gumbs, J. Appl. Phys. {\bf 107}, 103710 (2010).

\bibitem{huang2}D. H. Huang, S. K. Lyo and G. Gumbs, \prb {\bf 79}, 155308 (2009).

\bibitem{pair}S. K. Lyo and D. H. Huang, \prb {\bf 73}, 205336 (2006).

\bibitem{fertig}L. Brey and H. A. Fertig, \prb {\bf 73}, 235411(2006).

\bibitem{fn1}Such inter-valley scattering would require momentum transfer comparable with
the distance between $\mathbf{K}$ and $\mathbf{K}^\prime$ points.

\bibitem{heat}D. H. Huang, T. Apostolova, P. M. Alsing and D. A. Cardimona,
\prb {\bf 69}, 075214 (2004).

\bibitem{ando}T. Ando and T. Nakanishi, J. Phys. Soc. Japan, {\bf 67}, 1704, (1998).

\bibitem{wakabayashi}K. Wakabayashi, Y. Takane, M. Yamamoto and M. Sigrist, New J. Phys. {\bf 11}, 095016, (2009).

\bibitem{fertig1}L. Brey and H. A. Fertig, \prb {\bf 75}, 125434 (2007).

\bibitem{damp}S. K. Lyo and D. H. Huang, \prb {\bf 66}, 155307 (2002).

\bibitem{fn2}A distinct line must be drawn between equilibrium distribution function
$f_j^{(0)}$ in absence of applied electric field and
stationary solution of the transport equation $\lim\limits_{t\rightarrow 0}\,f_j(t)$.
\end{references}
\end{document}